\documentclass[10pt]{emulateapj}
\usepackage{apjfonts}
\usepackage{graphicx}
\usepackage{psfig}
\slugcomment{{\sc Submitted to ApJ:}}
\newcommand{\begit}{\begin{itemize}}
\newcommand{\enit}{\end{itemize}}
\newcommand{\begen}{\begin{enumerate}}
\newcommand{\enen}{\end{enumerate}}

\newcommand{\beq}{\begin{equation}}
\newcommand{\eeq}{\end{equation}}
\newcommand{\beqa}{\begin{eqnarray}} 
\newcommand{\eeqa}{\end{eqnarray}} 

\begin{document}

\title{Accelerating Compact Object Mergers in Triple
Systems with the Kozai Resonance:  A Mechanism for ``Prompt'' 
Type Ia Supernovae, Gamma-Ray Bursts, and Other Exotica}

\author{Todd A.~Thompson\altaffilmark{1}}
\affil{Department of Astronomy and
Center for Cosmology \& Astro-Particle Physics, \\
The Ohio State University, Columbus, Ohio 43210, USA;
thompson@astronomy.ohio-state.edu}
\altaffiltext{1}{Alfred P.~Sloan Fellow}

\begin{abstract}
White dwarf-white dwarf (WD-WD) and neutron star-neutron star (NS-NS) mergers may produce Type Ia supernovae and gamma-ray bursts (GRBs), respectively. A general problem is how to produce binaries with semi-major axes small enough to in significantly less than the Hubble time ($t_{\rm H}$), and thus accommodate the observation that these events closely follow episodes of star formation.  I explore the possibility that such systems are not binaries at all, but actually coeval, or dynamical formed, triple systems. The tertiary induces Kozai oscillations in the inner binary, driving it to high eccentricity, and reducing its gravitational wave (GW) merger timescale. This effect significantly increases the allowed range of binary period $P$ such that the merger time is $t_{\rm merge}<t_{\rm H}$. In principle, Chandrasekhar-mass binaries with $P\sim300$\,days can merge in $\lesssim t_{\rm H}$ if they contain a prograde solar-mass tertiary at high enough inclination.  For retrograde tertiaries, the maximum $P$ such that $t_{\rm merge}\lesssim t_{\rm H}$ is yet larger. In contrast, $P\lesssim0.3$\,days is required in the absence of a tertiary. I discuss implications of these findings for the production of transients formed via compact object binary mergers. Based on the statistics of solar-type binaries, I argue that many such binaries should be in triple systems affected by the Kozai resonance. If true, expectations for the mHz GW signal from individual sources, the diffuse background, and the foreground for GW experiments like {\it LISA} are modified. This work motivates future studies of triples systems of A, B,and O stars, and new types of searches for WD-WD binaries in triple systems.

\end{abstract}

\keywords{stars: binaries: close, gravitational waves --- celestial mechanics, stellar dynamics 
--- stars:neutron, white dwarfs --- supernovae: general}

\section{Introduction}
\label{section:introduction}

Stellar mass compact object mergers driven by gravitational
wave (GW) radiation may power many
types of observed astrophysical transients and produce rare
stellar exotica.  Notably, the merger of white dwarf-white dwarf (WD-WD) 
binaries has been suggested as a mechanism for producing Type Ia
supernovae (Iben \& Tutukov 1984; Webbink 1984), and neutron star-neutron star (NS-NS)
or NS-black hole (BH) mergers are a leading model for the 
central engine of short-duration gamma-ray bursts 
(GRBs) (Ruffert \& Janka 1999; Janka et al.~1999).  
Other types of compact object mergers such as
those of BH-Helium star or BH/NS-WD binaries have been identified as
possible mechanisms for long-duration GRBs 
(Fryer \& Woosley 1998, Fryer et al.~1999ab),
or peculiar supernovae (Metzger 2011).  
WD-WD mergers may also produce companionless millisecond 
pulsars (Saio \& Nomoto 1985) 
and R CrB stars (Webbink 1984, Nelemans et al.~2001).
Finally, close compact object binaries brought near enough 
for mass transfer power a variety of astrophysical
phenomena.  One example is AM CVn stars (Warner 1995), 
which may produce faint ``.Ia'' supernovae 
(Bildsten et al.~2007).

Estimates suggest that $\sim2$\% of the all the stars  
born with zero-age main sequence masses in the range 
$\sim2.5-8$\,M$_\odot$ become Ia supernovae
(e.g., Horiuchi \& Beacom 2010; but, see Maoz 2010).  If sub-Chandrasekhar mass 
binaries can contribute to the Ia rate (e.g., van Kerkwijk 
et al.~2010), the IMF implies that a still smaller fraction 
of all stars produce Ia's.  Similarly, the massive star birthrate,
which is approximately equal to the Type-II supernova rate, 
$\sim10^5$\,Gpc$^{-3}$ yr$^{-1}$ (e.g., Horiuchi et al.~2009),
is $\sim20-600$ times larger than the NS-NS merger
rate (Kalogera et al.~2001, 2004ab; O'Shaughnessy et al.~2008).

These comparisons imply that although a large fraction of all
stars are in binaries, only a small fraction have the requisite 
characteristics to produce compact object binaries 
that can merge via GW emission on a timescale short 
compared to the Hubble time.
This issue is particularly important since there is evidence
for a ``prompt'' subset of Ia supernovae that track the star formation
rates of galaxies (Scannapieco \& Bildsten 2005; Mannucci et al.~2006).
``Prompt'' may mean $\lesssim1$\,Gyr after a burst of star formation,
or potentially even $\lesssim200$\,Myr (Aubourg et al.~2008; 
Brandt et al.~2010; Maoz \& Badenes 2010; Maoz et al.~2010; Maoz et al.~2011).
Similarly, short-duration GRBs occur in
both quiescent and actively star-forming galaxies (e.g., Berger 2009).  
In order to solve this problem, most studies focus on binary evolution
channels that make more compact object binaries with small enough
semi-major axis that a merger via GW radiation occurs ``promptly''
(e.g., Belczy{\'n}ski \& Kalogera 2001; Ruiter et al.~2009).
Indeed, common envelope evolution of compact stellar binary systems
can lead to compact object binaries whose GW merger time is significantly 
less than the Hubble time (Iben \& Tutukov 1984, 1985, 1987).

In this paper I forward an alternate hypothesis: 
the rare binary systems that produce compact object
mergers are not binaries at all, but instead hierarchical
triple systems.  In such systems, Kozai (1962) showed that the
tertiary induces oscillations in the orbital eccentricity
of the inner binary via a secular resonance.  In the 
absence of general relativistic effects, the maximum 
eccentricity attained by the inner binary is
\beq
e_{\rm max}=\left(1-\frac{5}{3}\cos^2 i\right)^{1/2}
\label{emax}
\eeq
where $i$ is the inclination of the outer orbit relative
to the plane of the inner binary. 
Because the GW merger timescale is a very strong function of 
eccentricity ($t_{\rm GW}\propto(1-e^2)^{7/2}$; Peters 1964, eq.~\ref{tgw}), 
the addition of a tertiary at high inclination to a binary system can
decrease $t_{\rm GW}$ dramatically. Blaes et al.~(2002) (hereafter BLS02) 
showed this explicitly for the case of triple systems of super-massive BHs,
which would be formed by successive galaxy-galaxy  mergers 
during structure formation.  Miller \& Hamilton (2002) 
independently applied this same idea to four-body interactions 
(binary-binary scattering) involving stellar mass BH binaries in 
globular clusters to accelerate the assembly of intermediate mass 
BHs in these systems. 

Here, I explore the possibility that WD-WD, NS-WD, and NS-NS
mergers driven by GW radiation, of relevance particularly for 
Ia supernovae, GRBs, and
other transients, are accelerated by the presence of a 
tertiary at high inclination.  I consider both 
coeval triple systems and triples formed
by binary-single and binary-binary scatterings in dense 
stellar environments.  I show explicitly 
that the GW merger timescale for a subset of compact object
binaries in triple systems,
whether coeval or dynamically formed, is significantly
decreased from the expectation for the binary alone.
This effect dramatically increases the range of semi-major axes
for which a merger
will occur in a single Hubble time ($t_{\rm H}$).  Put another 
way, some systems which one might think have no hope of merging
in $t_{\rm H}$, actually merge in $\ll t_{\rm H}$ with a 
suitably placed tertiary.  This will affect the population 
synthesis of Ia- and GRB-producing compact object binaries, 
and may significantly affect the overall rate, even though
there are fewer triple systems than lone binaries.

These statements quickly raise the question of whether or not the 
parent population of triples that produce binary 
compact objects can accommodate the observed 
Ia or GRB rates.  Although a complete discussion of this issue is
beyond the scope of this paper, 
there are several reasons to believe that in the case of 
WD-WD mergers particularly, all such systems are 
(or were) triple. 
First, although only $\sim5-10$\% of all solar-type
stars are thought to be in triple systems (Tokovinin et al.~2006; Raghavan et al.~2010), 
fully $\sim50$\% and $\sim100$\% of all close solar-type binaries with orbital period 
$P\lesssim10$ and $\lesssim3$\,days, respectively, are in triple 
systems (Tokovinin et al.~2006; Pribulla \& Rucinski 2006).  
Second, Fabrycky \& Tremaine (2007) (hereafter FT07) have shown that 
the peak in the observed binary period distribution at $\sim3$\,days
can be accounted for by the combined action of tidal friction between 
the two stars in the inner binary and Kozai  oscillations
induced by a hierarchical tertiary (see also Mazeh \& Shaham 1979; 
Wu \& Murray 2003; Wu et al.~2007; Perets \& Fabrycky 2009).
Thus, the closest binaries --- those most likely to merge in less than a 
Hubble time --- are precisely those that are most likely
to be triples.  Additionally, although little is known 
about the triple fraction of the A/B stars that give rise to WDs sufficiently
massive to be plausible double-degenerate Ia progenitors
and the massive O/B stars ($\gtrsim10$\,M$_\odot$) that produce
NSs, there is evidence that the multiplicity of stars increases as a
function of stellar mass (Lada 2006; Raghavan et al.~2010).
Since the intermediate-mass  and massive stars that 
produce WD and NS binaries become Cepheids during their
post main-sequence evolution, the high 
occurrence of triples among such systems (Evans et al.~2005)
further motivates consideration of their eventual
compact objects.\footnote{Typically, 
a Cepheid with a $\simeq1$\,day period corresponds to a 
$\sim3$\,M$_\odot$ main sequence A0 star, which produces
a $\simeq0.7$\,M$_\odot$ WD (e.g., Kalirai et al.~2008), the minimum required for
a Chandrasekhar mass, equal mass ratio binary.}

The high percentage of close solar-type binaries containing tertiaries
motivates this paper in part.  The (albeit limited) statistics on such
systems implies that the rate of Ia supernovae and GRBs
compared to their overall parent population might be
accommodated by the triple fraction (see \S\ref{section:discussion}). 

There is an additional motivation for considering triple systems
as the progenitors for Ia supernovae.  The delay-time distribution
of Ia supernovae implies that many systems explode on 1-10 Gyr 
timescales (e.g., Totani et al.~2008), and this fact implies a certain number density of progenitor
systems in the Galaxy.  Simple estimates suggest that the nearest progenitor
system is of order just $\sim50$\,pc away. So far, there have been no 
confirmed identifications
of WD-WD binaries with combined mass greater than the Chandrasekhar mass
discovered (see Mullally et al.~2009 for a recent compilation).  
Although there are a number of potential explanations
for this fact --- e.g., sub-Chandrasekhar mass mergers or single-degenerate
systems dominate the Ia rate --- an additional possibility is that the reason
no such systems have been found is that they are in triple systems.
Since many such systems would be composed of a close WD-WD binary with a
hierarchical main sequence tertiary, such systems would not have been found
by searches that color-select WD binaries (Napiwotzki et al.~2001; 
Badenes et al.~2009; Brown et al.~2010).
Additionally, since the massive WD binaries that produce Ia supernovae 
may often have a lower-mass tertiary that evolves subsequently, triple
WD systems may appear as a single WD (the young, low mass tertiary),
and without a strong radial velocity signal from the distant,
dimmer, but more massive WD-WD binary.  Although a detailed discussion
of these effects on the selection of triple systems is beyond the scope
of this work, I consider the lack of 
observed WD-WD progenitors to be an additional motivation
for exploring the strong triple hypothesis that all Ia progenitors
are triple systems.

The remainder of this paper is organized as follows.
In \S\ref{section:scenarios}, I discuss the types
of systems that will be affected by Kozai oscillations.
Although I outline some of the evolutionary processes that 
will affect triple-star evolution (e.g., common envelope evolution
and mass loss), 
for the purposes of this paper, I simply assume that binary
compact objects are formed with a range of semi-major axes and masses, 
and with tertiaries with a variety of masses, semi-major axes,  
and inclinations relative to the inner binary.
I then calculate the evolution
of these representative systems using the methods described in  \S\ref{section:method}
and Appendix \ref{appendix:estimate}.  The results are presented
in \S\ref{section:results}.  In \S\ref{section:discussion},
I discuss the results and the implications for transients produced
from these mergers, and the overall rate.  I also discuss implications of 
these results for the gravitational wave foreground and background,
as well as the signal from individual sources,  for {\it LISA}
(see also Gould 2011).

\section{Scenarios}
\label{section:scenarios}

\subsection{WD-WD Binaries \& Triples}
\label{section:wdwd}

Consider a system of two $\sim2-8$\,M$_\odot$ (A5-B3) 
main sequence stars (e.g., most commonly
2+2, 3+2\,M$_\odot$, etc., and perhaps preferentially ``twins''; 
Pinsonneault \& Stanek 2006), $m_0$ and $m_1$,  
in a close binary with semi-major axis $a_1$,
and a coeval hierarchical tertiary of mass $m_2$
and semi-major axis $a_2$.  The minimum value of 
$a_2$ is set by stability of the triple system:
$a_2\gtrsim 3 a_1$ (Eggleton \& Kiseleva 1995; 
Mardling \& Aarseth 2001).

If $a_1$ is less than a few AU, then the binary will undergo one or more 
mass transfer and common envelope (CE) evolutionary phases, that decrease
$a_1$ from its initial value by a factor of $\sim10-100$ (Iben \& Tutukov
1984, 1985; Iben \& Livio 1993; Ruiter et al.~2009).  
Although the physics of the CE phase(s) is
uncertain, it is the primary mechanism for producing close 
WD-WD binaries that will merge in substantially less than 
$t_{\rm H}$ (see eq.~\ref{tgw}).

In the triple system considered, mass loss from the inner 
binary during the CE phase(s) will in general increase $a_2$ 
from its initial value, typically by a factor of 
$\sim2-10$, depending on the masses of system's constituents,
and the final WD masses of the inner binary.  Mass loss
from $m_2$ as it evolves to a WD will also increase $a_2$.  
If $a_2$ is less than a few AU either before or after the 
inner binary evolves, then there will be a complicated phase 
of triple CE evolution where the binary is engulfed by the tertiary,
potentially decreasing both $a_2$ and $a_1$.
A number of related evolutionary channels for triples 
were sketched by Iben \& Tutukov (1999).

In this paper, I show that for a range of parameters the timescale
for the merger of WD-WD binaries via GWs is dramatically decreased
by the presence of a tertiary.  As implied by equation (\ref{emax}),
and as discussed in detail in Sections \ref{section:method} and 
\ref{section:results}, this mechanism requires (1) that the tertiary
be at high relative inclination with respect to the inner binary
and (2) that $a_2/a_1$ must be in the range of $\sim3-100$.  
Both are expected to be modified by stellar evolution 
preceding WD formation.  The latter 
is important since the combination of mass loss and 
$a_1$ evolution during the CE phase of a close binary will in general increase
$a_2/a_1$ from its initial value by a factor of $20-1000$.
However, the phase of triple CE evolution that may occur
as the tertiary evolves can in principle decrease $a_2$ and 
$a_1$, and increase or decrease $a_2/a_1$.
The evolution in inclination is also important
since equation (\ref{emax}) and the fact that $t_{\rm GW}\propto(1-e^2)^{7/2}$
imply that $t_{\rm GW}$ will be a very strong function of $i$.
For a random distribution of inclination angles
between $0^\circ$ and $90^\circ$ (prograde tertiaries),
the probability of having 
$i>70^\circ$, 80$^\circ$, 85$^\circ$, and 89$^\circ$
is $\cos i\simeq0.3$, 0.2, 0.09, and 0.02, respectively.
Although the distribution of tertiary inclinations is 
unknown, the results of FT07 imply that if a uniform distribution 
in $\cos i$ is assumed when the stars first form, the resulting 
distribution, after many Kozai times (eq.~\ref{tk}), is quite flat
(their Fig.~7), but with peaks near the critical Kozai angles
($i\sim39^\circ$ and $141^\circ$).  Because tides detune the Kozai
resonance, if the tertiary is at high $i$, the binary's
semi-major axis and eccentricity will be affected by the combination of 
Kozai oscillations and tidal friction before 
WD formation  (FT07).    In particular, the inner binary will be driven to lower eccentricity
than if Kozai acted alone (eq.~\ref{emax}).  If there is a
CE evolutionary phase as the stars evolve, and mass transfer, the inner binary may 
well be nearly circular at the time of WD formation since the Kozai 
resonance will be terminated by such strong interaction between the
inner components.  However, subsequent to this complicated
evolution, one expects Kozai to again operate after compact 
object formation.  The system will then be driven to high eccentricity, 
and rapidly coalesce
via GW radiation.

Given the complications posed by 
triple star evolution, and in particular triple CE evolution,
I simply assume that such systems lead to binary 
WDs with a range of $a_1$, together with a tertiary WD or main 
sequence star with a range of inclinations and $a_2$. 
The combination of triple star evolution with dynamics,
mass transfer, and CE evolution, together with population
synthesis is crucial from a theoretical perspective
for evaluating the mechanism proposed
here for rapid coalescence of WDs via gravity waves.
However, these theoretical complications aside, 
the results of Tokovinin et al.~(2006) and Pribulla \& Rucinski (2006) strongly 
motivate an observational search for WD-WD binaries with main sequence
tertiaries.  The discovery and characterization of such
systems --- in particular the quantities $i$ and $a_2/a_1$ ---
will be the ultimate arbiter in 
establishing the importance of the Kozai mechanism 
for rapid compact object mergers.

Although I have made no attempt to model specific systems,
recent work on the exotic planetary nebula SuWt 2 provides 
further motivation since it is thought to contain 
a close A-star binary with a (possibly WD) $\sim0.7$\,M$_\odot$
tertiary (Exter et al.~2010).  One can also imagine
an important role for the Kozai mechanism in driving 
binaries composed of a WD and a main sequence star to contact,
as in the single-degenerate scenario for Ia supernovae
(Whelan \& Iben 1973).

Similar scenarios with different mass components may produce 
AM CVn or R CrB stars, depending on the stability of mass transfer
as the WDs interact (Webbink 1984, Nelemans et al.~2001),
or millisecond pulsars (Saio \& Nomoto 1985).

In globular clusters, close WD-WD binaries may pick up a 
tertiary from the dense stellar field via either binary-single
or binary-binary scattering.  The latter should dominate the rate.
This mechanism for the formation of triple systems that lead to 
Kozai-induced mass transfer has been studied by 
Ivanova (2008) and Ivanova et al.~(2008,2010), and has been found
to be important for producing the observed X-ray binary populations
of globular clusters (see also Fregeau et al.~2004, 2009).   
As I show in \S\ref{section:results},
if the tertiary is captured into a high-inclination prograde orbit, 
or a retrograde orbit with $i\lesssim110-120^\circ$, the GW merger 
timescale for the inner binary can be very short.  Since such systems
will have very high eccentricity, their GW signals will be peaked
at high frequencies (\S\ref{section:results}), and for individual
systems {\it LISA} will see GW ``pulses'' at periastron (Gould 2011).

\subsection{BH, NS-BH, NS, WD Scenarios}
\label{section:ns_scenario}

For binaries with a BH or NS in triple systems, the supernova
explosions that accompany NS  (and possibly BH) formation may in some cases unbind
the inner binary or the tertiary, depending on the binary mass ratio, the 
tertiary mass, the eccentricities of the inner and outer orbits, 
and the magnitude and direction of the ``kick'' velocity 
given to the NS at birth.  For NS-NS-producing binaries of relevance
for short-duration GRBs, one imagines an inner binary with 
components of $\sim9-12$\,M$_\odot$ (e.g, 9+10, 9+9, 11+10\,M$_\odot$, etc.),
forming via standard scenarios (Bhattacharya \& van den Heuvel  1991),
and with a relatively distant tertiary with $m_2\sim5-8$\,M$_\odot$, which
may be either a main-sequence B star, or perhaps a BH formed by
a more massive star.  Depending on the mass of the system expelled
before and during the supernovae that produce the NSs and the magnitude
and direction of their kicks, the tertiary may remain bound,
but with small $a_2$ preferred (see, e.g., Fig.~10 of 
Kalogera 1996 for the binary case; Hills 1983).

After the supernovae, 
the resulting NS-NS binary will be subject to Kozai oscillations
by the tertiary, depending on its semi-major
axis and inclination.  If the tertiary is a main-sequence star,
it will become a massive WD with $\sim1-1.4$\,M$_\odot$.  As in 
the WD-WD case, because the timescale for mass loss  
during this transformation is long compared to the orbital
times in the system, it should remain bound, albeit with 
larger $a_2/a_1$.  In addition, the system may also undergo 
triple CE evolution as the tertiary evolves.  After this is 
complete, if the system persists, Kozai cycles will then resume. 

Many similar scenarios can be considered for NS-WD binaries or 
BH-NS/WD binaries in coeval triple systems,
although the survival of the triple through the formation of the compact
object binary must be evaluated case-by-case.
A discussion of one such system in the context of 
Kozai oscillations in a hierarchical triple system is 
given in Champion et al.~(2008) and Freire et al.~(2011) 
for the system PSR J1903+0327.

As discussed in \S\ref{section:wdwd} in the WD-WD case, 
NS-NS/WD/BH binaries that undergo Kozai oscillations
can also be formed dynamically in dense stellar environments
via binary-single and binary-binary scattering.
A significant fraction of such interactions would be expected to 
produce a stable triple system (see MH02; Wen 2003;
Ivanova 2008; Ivanova et al.~2008, 2010).

%%%%%%%%%%%%%%%%%%%%%%%%%
\begin{figure*}
\centerline{\includegraphics[width=8.5cm]{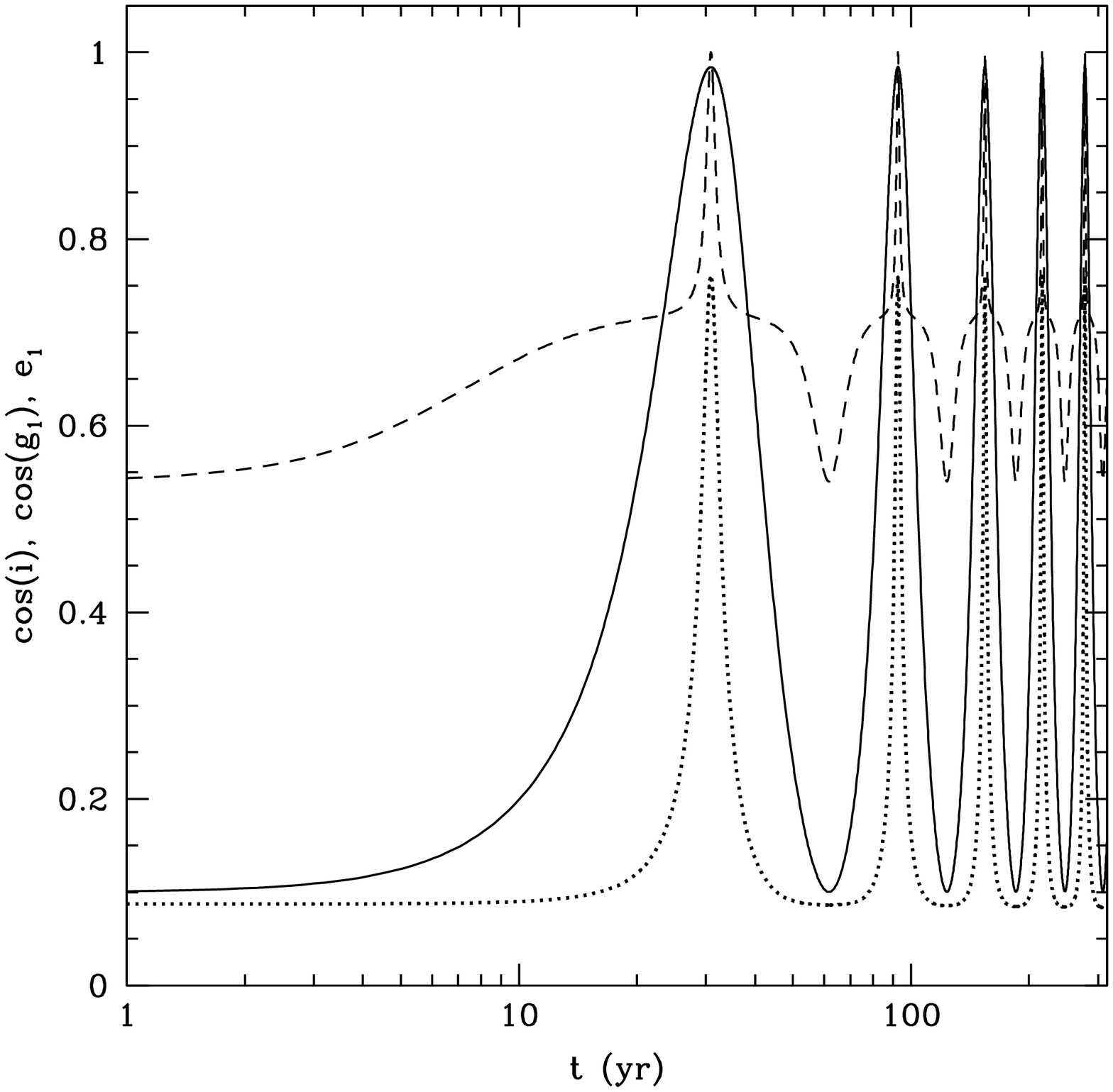}
\includegraphics[width=8.5cm]{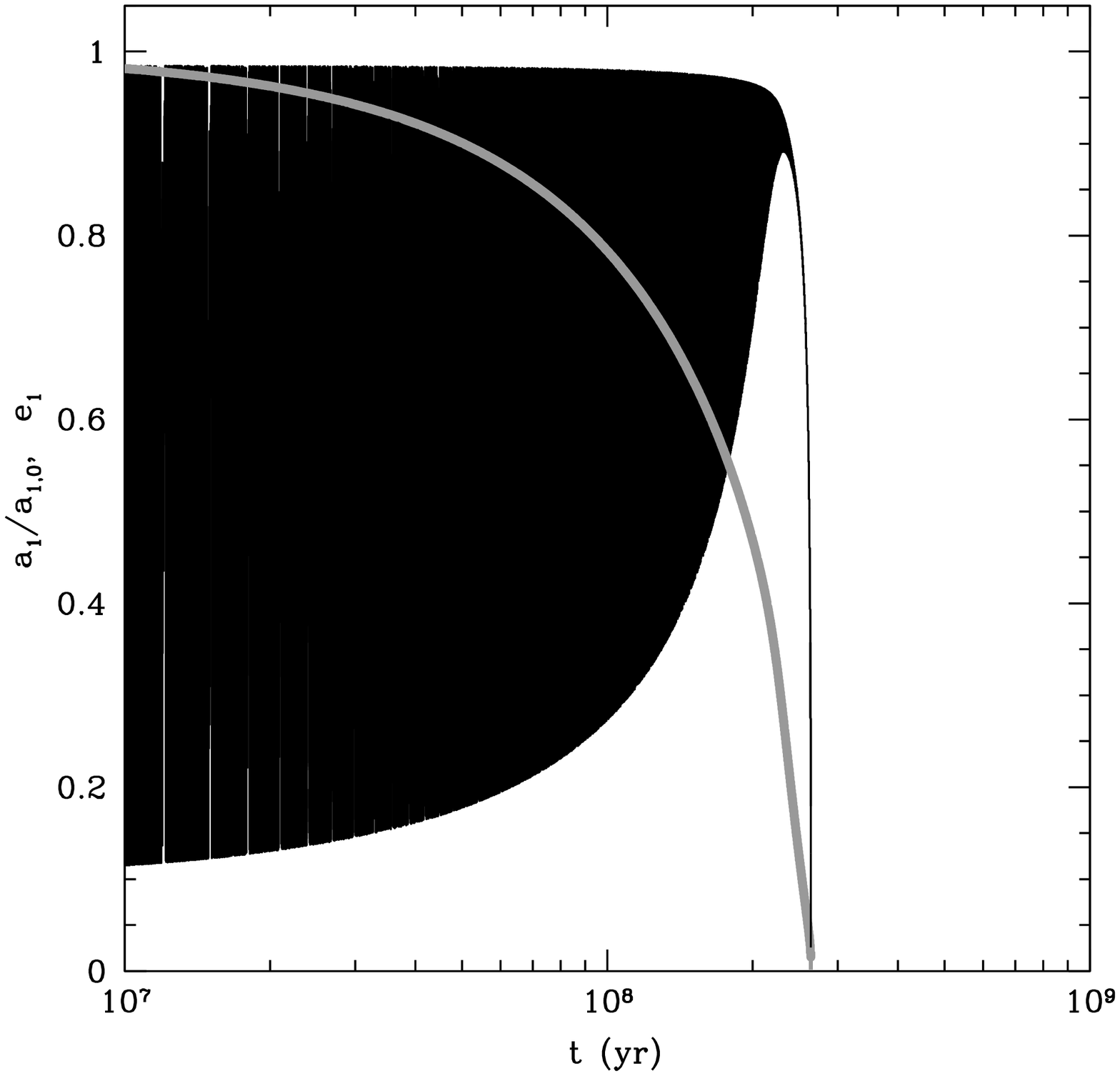}}
\caption{Time evolution of a triple system with 
$a_{1,0}=0.05$\,AU and $a_{2,0}/a_{1,0}=20$, with initial
conditions $i_0=85^\circ$, $e_{1,0}=e_{2,0}=0.1$, 
$g_{1,0}=0^\circ$, $g_{2,0}=90^\circ$, $m_0=0.8$\,M$_\odot$,
$m_1=0.6$\,M$_\odot$, and $m_3=1.0$\,M$_\odot$ (i.e., 
a WD-WD binary with a main sequence companion).
The nominal GW merger timescale without the tertiary is 
$\simeq3\times10^{12}$\,yr, but with the tertiary, 
$t_{\rm merge}\simeq2.5\times10^{8}$\,yr. {\it Left panel:}
$\cos(i)$ (dotted), $\cos(g_1)$ (dashed), $e_1$ (solid).
{\it Right panel:} $e_1$ (black), $a_1/a_{1,0}$ (gray).
}
\label{fig:t}
\end{figure*}
%%%%%%%%%%%%%%%%%%%%%%%%% 

\section{Method}
\label{section:method}

\subsection{Timescales}

I consider an inner compact object binary with semi-major axis 
$a_1$, eccentricity $e_1$, argument of periastron $g_1$, and masses
$m_0$ and $m_1$.  The hierarchical tertiary has mass $m_2$,
semi-major axis $a_2$, argument of periastron $g_2$,  and 
mutual inclination with respect to the inner binary of $i$.
Subscripts of ``0'' are used to indicate initial values
(e.g., $e_{1,0}$).

There is a strong hierarchy of timescales in the problem of triple systems
consisting of a compact object binary.  First,
in the limit of high eccentricity, the GW merger timescale of 
the inner binary is (Peters 1964)
\beqa
t_{\rm GW}&=&
\frac{3}{85}\frac{a_1}{c}\left(\frac{a_1^3c^6}{G^3m_0m_1M}\right)(1-e_1^2)^{7/2} \nonumber \\
&\simeq&1.6\times10^{13}{\rm \,\,yr}\,\,
\left(\frac{2M_\odot^3}{m_0m_1M}\right)\left(\frac{a_1}{0.1{\rm AU}}\right)^4(1-e_1^2)^{7/2},
\label{tgw}
\eeqa
where $M=m_0+m_1$.  The low-eccentricity version of 
equation (\ref{tgw}) is $t_{\rm GW}\times(425/768)(1-e_1^2)^{-7/2}$ (Peters 1964).
Throughout this paper I focus on stellar-mass binaries whose 
initial semi-major axis is large enough that the 
nominal value of $t_{\rm GW}$, in the absence of a tertiary,
is greater than the Hubble time, $t_{\rm H}\simeq14$\,Gyr.  This implies 
values for $a_{1,0}$ larger than $\sim0.017$\,AU, depending on 
the masses of the binary components considered.

As discussed by BLS02 and MH02 
in the context of supermassive BH mergers 
and the formation of intermediate mass BHs, respectively, 
GR periastron precession ``de-tunes'' the secular Kozai resonance.
This effect in general decreases the maximum eccentricity attainable 
at fixed tertiary inclination (see Appendix of BLS02 and the discussion 
in FT07).  The timescale (period) for GR precession (e.g., eq.~23 of FT07) is
\beqa
t_{\rm GRp}&=&
\frac{1}{3}\frac{a_1}{c}\left(\frac{a_1 c^2}{GM}\right)^{3/2}\hspace*{-0.2cm}(1-e_1^2) \nonumber \\
&\simeq&3.7\times10^4{\rm \,\,yr}\,\,
\left(\frac{2M_\odot}{M}\right)^{3/2}\left(\frac{a_1}{0.1{\rm AU}}\right)^{5/2}(1-e_1^2).
\label{tgrp}
\eeqa
and the Kozai timescale is (Innanen et al.~1997; Holman et al.~1997)
\beqa
t_{\rm K}&=&\frac{4}{3}\left(\frac{a_1^3 M}{Gm_2^2}\right)^{1/2}\left(\frac{b_2}{a_1}\right)^3   \nonumber \\
&\simeq& 77{\rm \,\,yr}\,\,
\left(\frac{a_1}{0.1{\rm AU}}\right)^{\hspace*{-0.1cm}3/2}
\hspace*{-0.1cm}\left(\frac{M}{2M_\odot}\right)^{\hspace*{-0.1cm}1/2}
\hspace*{-0.1cm}\left(\frac{M_\odot}{m_2}\right)
\hspace*{-0.1cm}\left(\frac{b_2/a_1}{20}\right)^{\hspace*{-0.1cm}3},
\label{tk}
\eeqa
where $b_2=a_2(1-e_2^2)^{1/2}$.  As discussed in BLS02, 
Kozai oscillations only operate if $t_{\rm K}<t_{\rm GRp}$.  
The strong dependence of $t_{\rm K}$ on $a_2/a_1$ implies
that there is a maximum $a_2$, beyond which Kozai oscillations
are ineffective.  Additionally, since 
$$t_{\rm GRp}/t_{\rm K}\propto a_1^4,$$
as the binary evolves to smaller $a_1$ because of GW radiation,
$t_{\rm K}$ eventually becomes larger than $t_{\rm GRp}$.
This can cause the binary to circularize before coalescence,
and it is the competition between $t_{\rm GRp}$ and $t_{\rm K}$
that determines much of the time evolution
of the system as it evolves towards merger.

Momentarily ignoring the complications of GR precession, one
can combine equation (\ref{emax}) with equation (\ref{tgw})
to get a rough order-of-magnitude sense of the importance of 
Kozai oscillations in triple systems for the rapid merger of
compact objects.  Taking the maximum eccentricity the system
reaches to be $e_{\rm max}$, the roughest estimate of the merger time
is simply
\beq
t_{\rm merge}\sim t_{\rm GW}(a_1,e_{\rm max})(1-e_{\rm max}^2)^{-1/2},
\eeq
where the factor $(1-e_{\rm max}^2)^{-1/2}$ corrects for the small
relative amount of time the system spends at high eccentricity (e.g., MH02).
Substituting, one finds that 
\beqa
t_{\rm merge}&\sim&
\frac{25}{153}\frac{a_1}{c}\left(\frac{a_1^3c^6}{G^3m_0m_1M}\right)\cos^6i\nonumber \\
&\sim&8.7\times10^9\,{\rm yr}\,\,
\left(\frac{2M_\odot^3}{m_0m_1M}\right)
\left(\frac{a_1}{0.1\,{\rm AU}}\right)^4
\left(\frac{\cos i}{0.2}\right)^6,
\label{est6}
\eeqa
which shows the very strong expected dependence on the inclination
angle.  The estimate of equation (\ref{est6}) is only valid for angles
in the critical Kozai range between $39^\circ\lesssim i\lesssim141^\circ$, and 
fails to account for important corrections to $e_{\rm max}$ from GR precession
(BLS02, MH02, Wen 2003, FT07).  In particular, it has
no dependence on $a_2/a_1$,  and its dependence on $a_1$ is 
only approximate.  In some regions of parameter space equation (\ref{est6})
thus grossly underestimates $t_{\rm merge}$ (see \S\ref{section:results}).  A much more
accurate, but less simply stated,  estimate of $t_{\rm merge}$ can be made using the method of Wen (2003)
that is accurate to a factor of a few over many decades in  
$t_{\rm merge}$ (see Appendix \ref{appendix:estimate}).

%%%%%%%%%%%%%%%%%%%%%%%%%
\begin{figure*}
\centerline{\includegraphics[width=8.5cm]{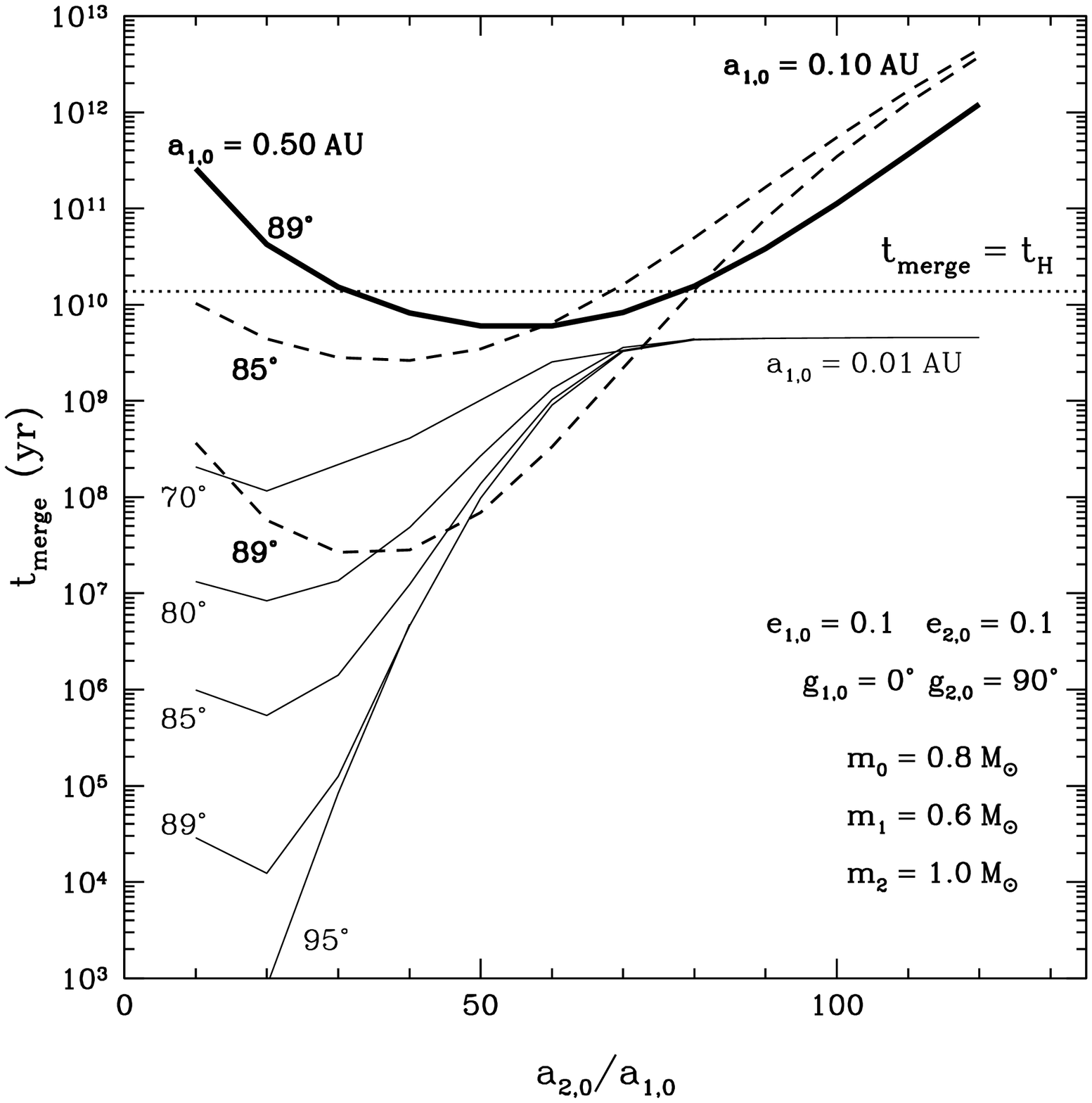}
\includegraphics[width=8.5cm]{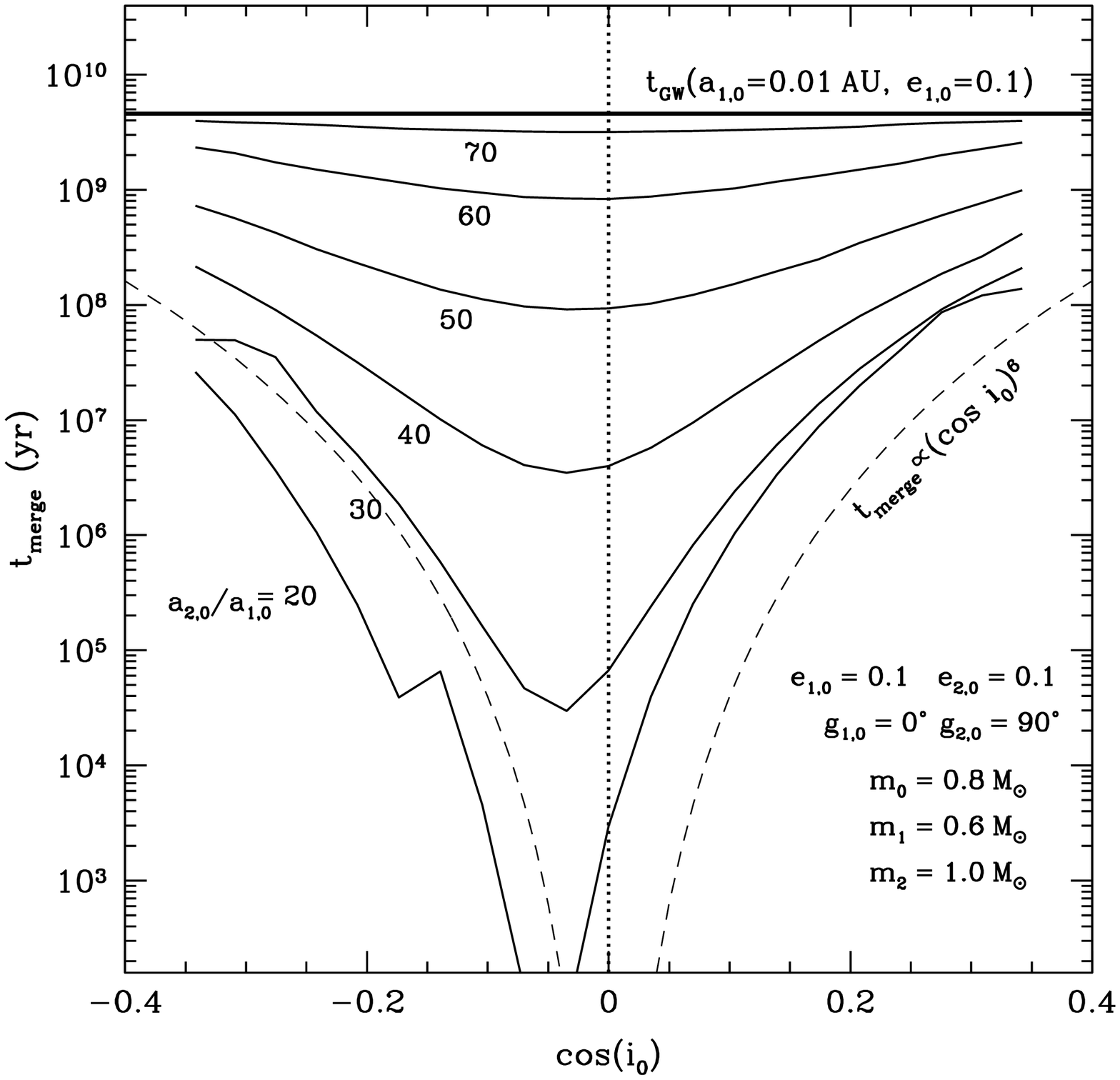}}
\caption{{\it Left Panel:} Merger time $t_{\rm merge}$ 
versus initial $a_{2,0}/a_{1,0}$, for 
$a_{1,0}=0.01$ (light solid), 0.1 (dashed), and 0.5\,AU (heavy solid),
for several values of $i_0$.
For sufficiently large $a_{2,0}/a_{1,0}$, the presence of the tertiary
does not speed up the merger and $t_{\rm merge}\rightarrow 
t_{\rm GW}(a_{1,0},e_{1,0}=0.1)$. 
{\it Right Panel:} $t_{\rm merge}$ versus $\cos(i_0)$ for 
$a_{1,0}=0.01$\,AU, and $a_{2,0}/a_{1,0}=20$, 30, 40, 50, 60, 70.  Note that 
for retrograde tertiaries $t_{\rm merge}$ becomes very short,
and very high eccentricity WD-WD interactions are possible.  
The scaling of $t_{\rm merge}\propto\cos^6i$ from equation 
(\ref{est6}) is shown for comparison.}
\label{fig:wd}
\end{figure*}
%%%%%%%%%%%%%%%%%%%%%%%%% 

\subsection{Equations, Assumptions, \& The Merger Time}
\label{section:numerics}

I solve the octopole-order equations for the secular
evolution of the orbital elements of the system, as given 
in BLS02 (their eqs.~11-17), 
which are based on the expressions derived in Ford et al.~(2000, 2004) 
(Krymolowski \& Mazeh 1999; Marchal 1990).  These equations
include both GR periastron precession 
and GW radiation (MH02; Wen 2003), but 
neglect tidal forces and treat the masses as point particles.
This amounts in part to neglecting
terms in the equation for the time evolution of the 
longitude of periastron of the inner binary associated
with both tidal and rotational bulges.
Both can suppress Kozai oscillations in a manner similar
to GR precession (eq.~\ref{tgrp}).   In particular, the
timescale associated with the apsidal motion induced by
a tidal bulge is (e.g., FT07)
\beqa
t_{\rm tide}&=&\frac{4}{15k}\left(\frac{a_1}{R}\right)^5
\left(\frac{a_1^{3}}{GM}\right)^{1/2}
\frac{(1-e_1^2)^5}{8+12e_1^2+e_1^4} \nonumber \\
&\simeq&2.35\times10^{15}{\rm \,yr}\,\,\left(\frac{a_1}{0.1\,{\rm AU}}\right)^{13/2}
\left(\frac{5000\,{\rm km}}{R}\right)^5\left(\frac{2\,{\rm M_\odot}}{M}\right)^{1/2} \nonumber \\
&\,&\hspace*{4cm}\times\frac{(1-e_1^2)^5}{8+12e_1^2+e_1^4},
\label{ttide}
\eeqa
where 
$R$ is the radius of the compact objects (assumed equal), 
$m_0=m_1$ has been assumed, and $k$ is the classical apsidal motion constant
(for the numerical estimate $k=0.1$).  
For sufficiently high eccentricity (e.g., $e_1\gtrsim0.9986$ for the parameters
of eq.~\ref{ttide}),  $t_{\rm tide}$ becomes less than $t_{\rm GRp}$, and one 
expects tides to become important to the evolution.  In some cases
the eccentricity can become large enough that periapsis approaches the 
WD radius.  In such cases, one expects strong tidal heating and circularization.
Although these effects are of interest in their own right, and
they might be especially important for the resulting transients
from such mergers, they are not captured in the current study,
and will be the subject of a future work.  For the present purposes,
it is sufficient to note that GW merger timescale on a scale
comparable to the compact object radius (whether NS or WD),
is ultrashort compared to the merger time of circular
binaries without a tertiary at the semi-major axes of 
interest ($>t_{\rm H}$).  Thus, if the binary does 
circularize on such scales, the merger timescale $t_{\rm merge}$
as estimated by numerical solution of the time-dependent equations,
is not dramatically affected.

While the merger time $t_{\rm merge}$ reported for binaries
calculated in \S\ref{section:results} is the time required
for the semi-major axis to reach the radius of the 
compact object $R_{\rm CO}$ 
(I take $R_{\rm WD}=5000$\,km and $R_{\rm NS}=10$\,km),
in some extreme cases
the periapsis of the orbit reaches $\sim R_{\rm CO}$.  In such cases,
I explicitly note that such solutions will be strongly affected 
by tides. The combined action of tidal
friction and Kozai oscillations on the period distribution of solar-type
binaries in triple systems has been considered in detail by 
 FT07 (see also Mazeh \& Shaham 1979; Kiseleva et al.~1998; Eggleton \& Kiseleva-Eggleton 2001;
Wu \& Murray 2003; Wu et al.~2007; Perets \& Fabrycky 2009).
A similar calculation of white dwarf and 
neutron star binaries in triple systems 
is saved for a future paper.

Finally, all of the initial configurations described below are
chosen to be stable when compared to the empirical 3-body
stability criterion of Mardling \& Aarseth (2001) (see
also Eggleton \& Kiseleva 1995).
For a discussion in the context of Kozai oscillations,
see BLS02.

\subsection{Numerics}

Standard methods (e.g., Bulrisch-Stoer; Press et al.~1992) 
are employed to solve the system of equations,
and I have done several checks to ensure the fidelity of the
results presented.  First, I have  
varied the numerical tolerance of the algorithm systematically
and found that the results presented here are converged.
As an additional check, I have verified that as $a_2$ becomes 
large and $t_{\rm K}\gg t_{\rm GRp}$, the solution for 
$t_{\rm merge}$ of the inner binary approaches 
the result of Peters (1964) for an isolated
binary.  Finally, I have spot checked my 
calculations directly against the 
code of BLS02 and the numerical results of Wen (2003),  
and find excellent agreement.

Some of the results presented make use of the 
simple approximate method of Wen (2003), 
described in Appendix \ref{appendix:estimate}.
Comparisons between this approximation 
and the actual solution of the time-dependent
problem are provided there.

%%%%%%%%%%%%%%%%%%%%%%%%%
\begin{figure*}
\centerline{\includegraphics[width=8.5cm]{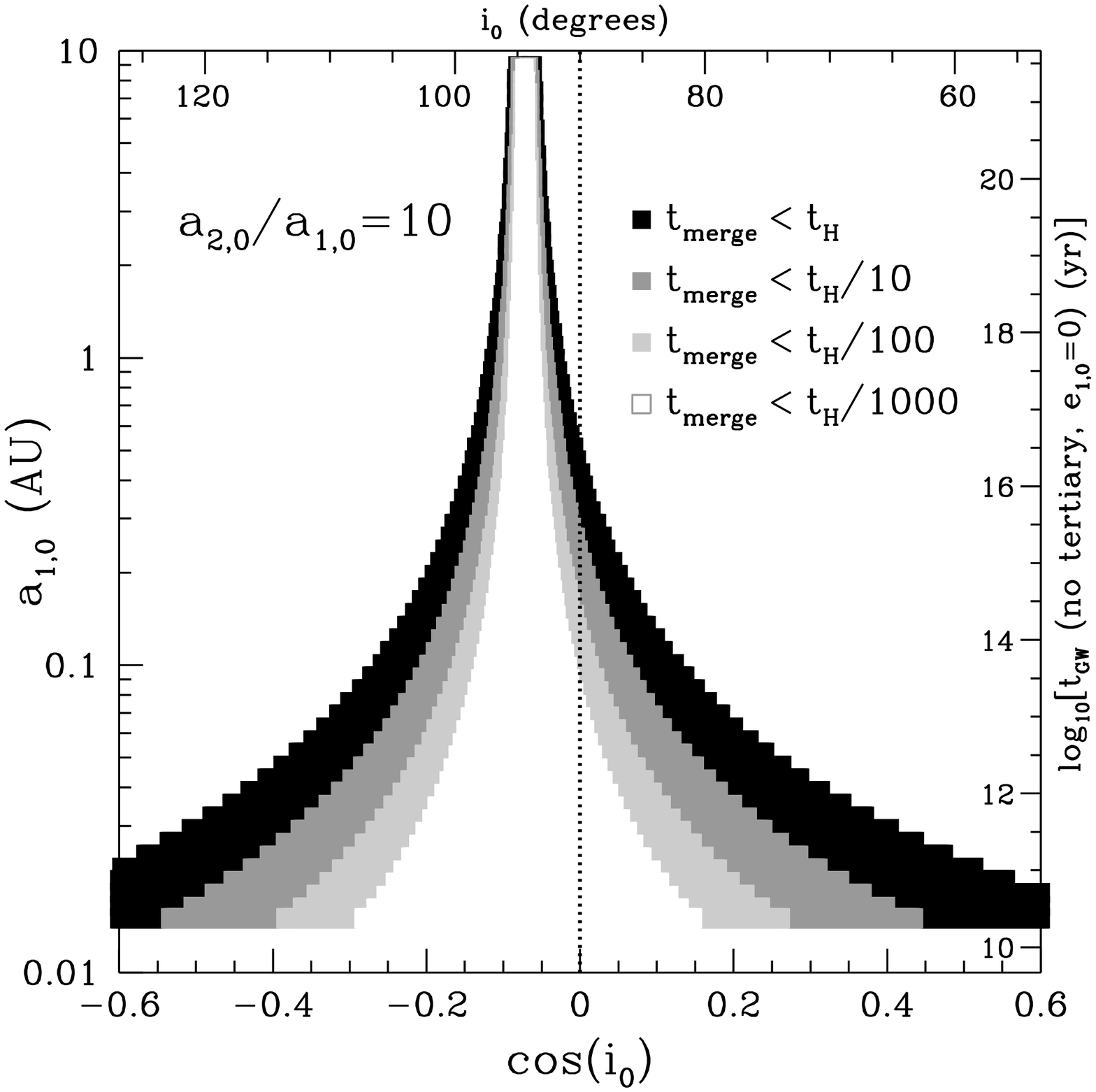}
\includegraphics[width=8.5cm]{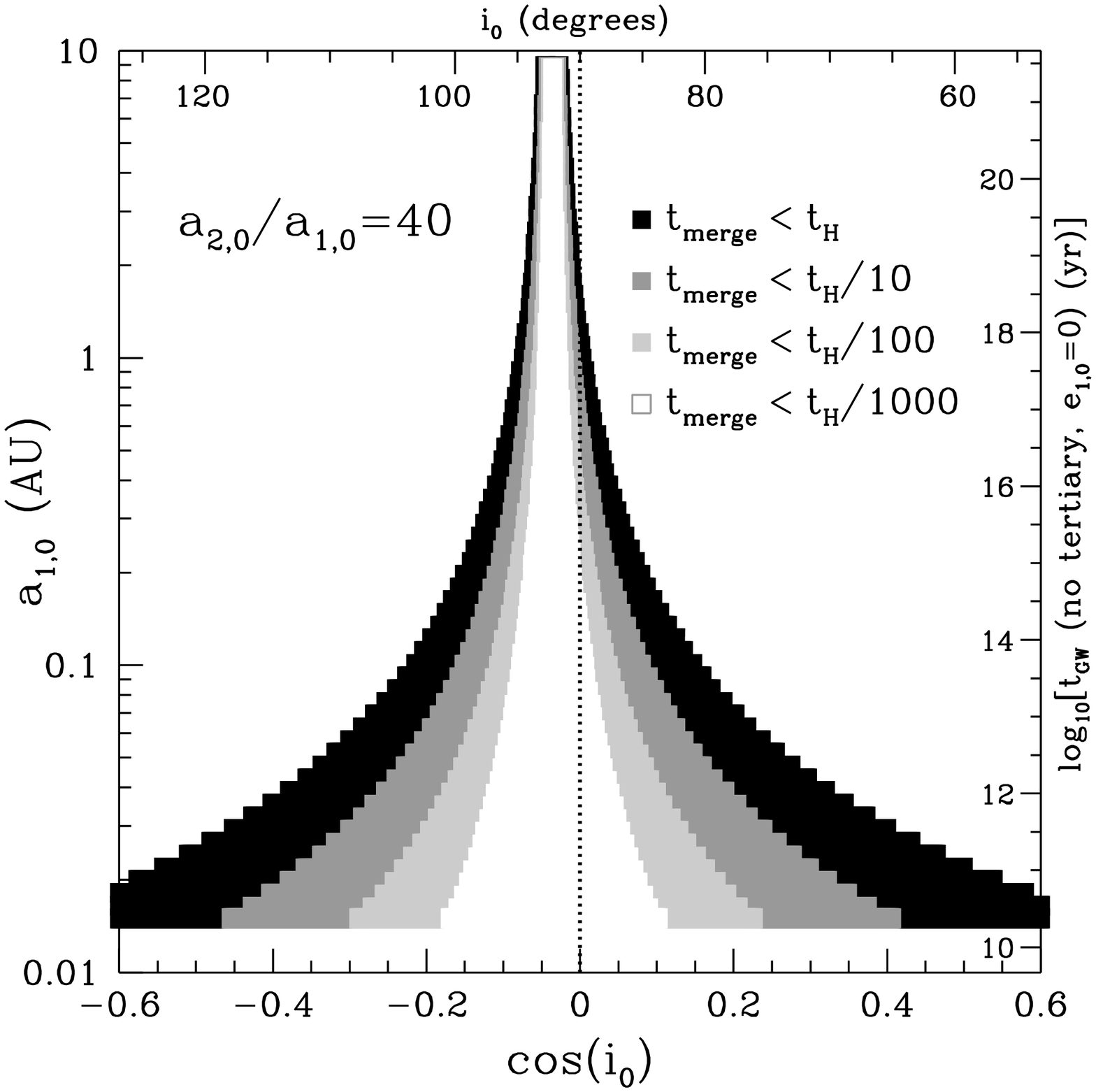}}
\centerline{
\includegraphics[width=8.5cm]{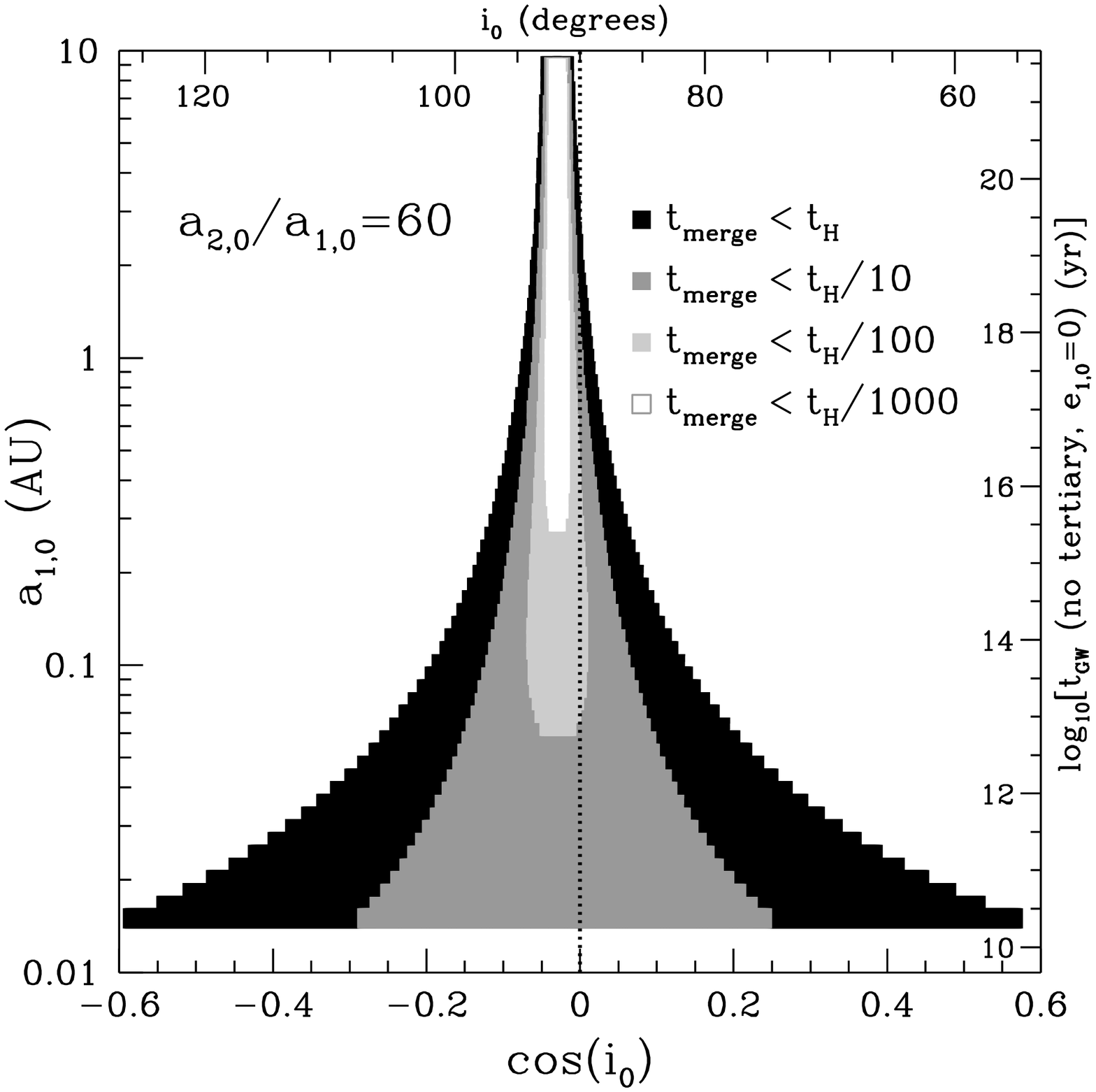}
\includegraphics[width=8.5cm]{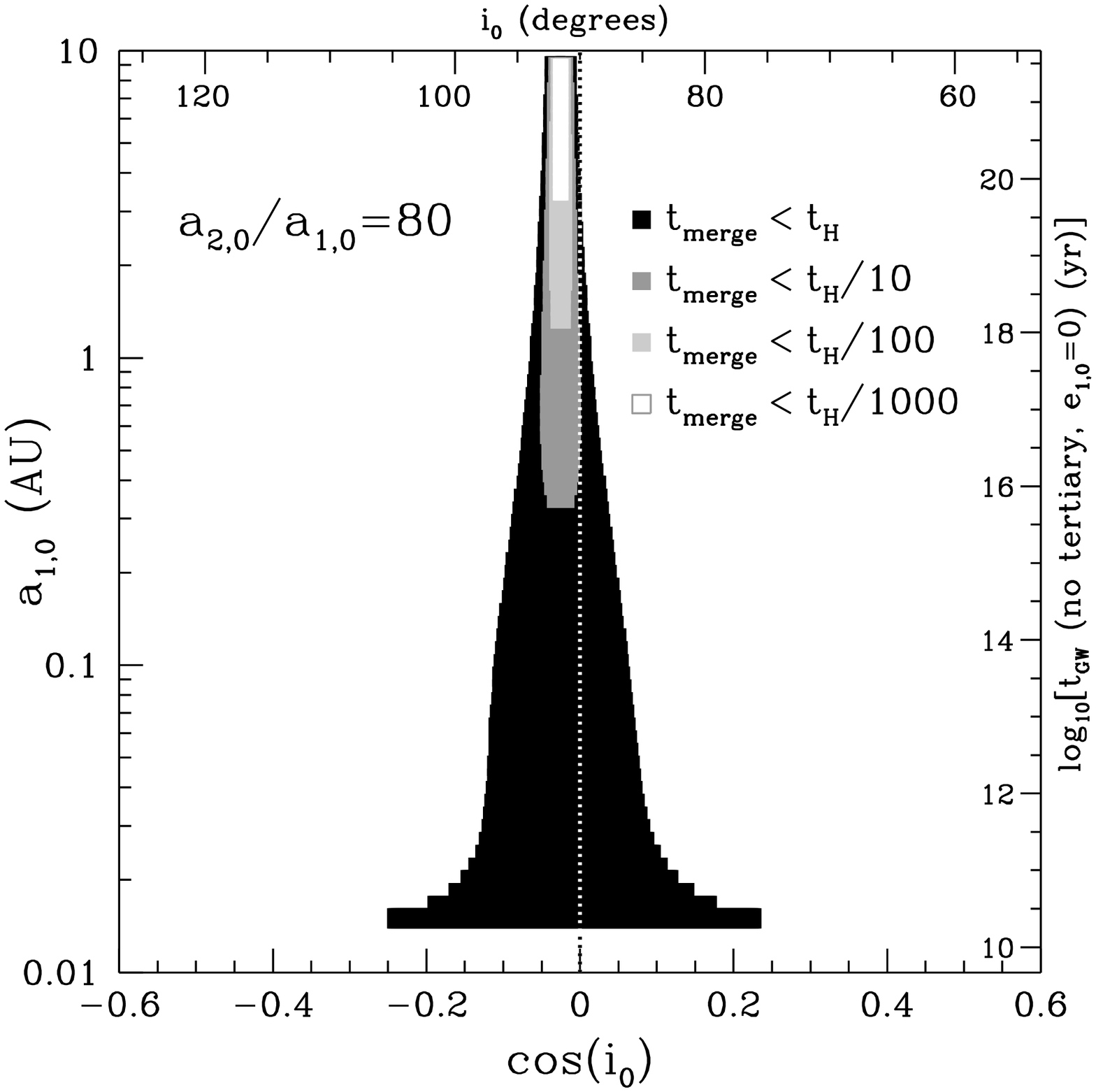}}
\caption{Range of allowed initial inner binary semi-major axis $a_{1,0}$
versus initial mutual inclination $i_0$, such that the inner binary
merges in $t_{\rm H}$, $t_{\rm H}/10$, $t_{\rm H}/100$, and $t_{\rm H}/1000$ 
(darkest to lightest), for outer tertiary semi-major axis $a_{2,0}/a_{1,0}
=10$ (top left), 40 (top right), 60 (bottom left), and 80 (bottom right),
computed by estimating $t_{\rm merge}$ using the algorithm
described in Appendix \ref{appendix:estimate} for 
$m_0=m_1=0.7$, $m_2=1.0$, $e_{1,0}=e_{2,0}=0.1$, 
$g_{1,0}=0^\circ$, and $g_{2,0}=90^\circ$.  Note that for 
these inner binary masses, only those with $a_{1,0}\lesssim0.015$\,AU
(off the bottom of the range shown) 
would be expected to merge in $t_{\rm H}$ in the absence of a
tertiary companion.  Note further that for tertiaries randomly 
distributed in $i$, $\cos(i)$ is proportional to the probability
of having such a system.}
\label{fig:ain}
\end{figure*}
%%%%%%%%%%%%%%%%%%%%%%%%%  

\section{Results}
\label{section:results}

The parameter space of possible masses and orbits is very large.
To restrict the total model space, I take the initial values of 
the orbital eccentricities and arguments of periastron to be 
$e_{1,0}=e_{2,0}=0.1$, $g_{1,0}=0^\circ$, and $g_{2,0}=90^\circ$ throughout
this paper.  
One expects the assumption of low initial eccentricity to 
be reasonable except in two cases of particular interest: 
(1) in dynamically formed triple
systems the eccentricity of the tertiary may be large $e_2\sim0.9$ (Ivanova 2008), 
and (2) in NS-NS binaries the NS kicks at birth may 
cause $e_1$ to be large.  Higher $e_1$ and $e_2$
generically lead to faster mergers, and thus the assumption 
of low initial eccentricities is conservative.

Figure \ref{fig:t} shows the time evolution of a selection of orbital
elements for the fiducial WD-WD case with $m_0=0.8$\,M$_\odot$,
$m_1=0.6$\,M$_\odot$, and $m_2=1.0$\,M$_\odot$.  The masses of $m_0$
and $m_1$ are chosen to sum to 1.4\,M$_\odot$ for illustrative purposes,
and $m_0\ne m_1$ so that the
octopole-order terms in the dynamical equations operate.  
Results for  $m_0=m_1=0.7$\,M$_\odot$ are not significantly different. 
The initial semi-major axis ($a_{1,0}=0.05$\,AU) 
is chosen so that the nominal GW merger timescale
in the absence of the tertiary is $\sim200 t_{\rm H}$.
The initial mutual inclination of the system is $i=85^\circ$
($\cos i\simeq0.09$). The left panel shows the early time 
evolution of $i$ (dotted), $g_1$ (dashed), and $e_1$ (solid).
The periodic changes in $e_1$ correspond to $t_{\rm K}$ (eq.~\ref{tk}).
The right panel shows the late-time evolution of $e_1$ and 
the semi-major axis, scaled by its initial value ($a_1/a_{1,0}$).
The system merges in $\simeq2.6\times10^8$\,yr, approximately 
$10^4$ times faster than without the tertiary.  At no time before
the very end of the calculation does the periapsis of the orbit 
reach $2\times R_{\rm WD}$. 

The results of many such calculations are presented in Figure \ref{fig:wd}.
The left panel shows $t_{\rm merge}$ as a function
of the initial value of $a_{2,0}/a_{1,0}$, for 
$a_{1,0}=0.01$\,AU (thin solid), 0.1\,AU (dashed), and 0.5\,AU (heavy solid),
for several different values of $i_0$ as labeled.  
The nominal (no tertiary) binary merger timescales via
GW radiation are $\sim5\times10^9$\,yr ($\sim0.3$\,$t_{\rm H}$), 
$\sim5\times10^{13}$\,yr ($3\times10^3$\,$t_{\rm H}$), and 
$\sim3\times10^{16}$\,yr ($2\times10^6$\,$t_{\rm H}$), for 
$a_{1,0}=0.01$\,AU, 0.1\,AU, and 0.5\,AU, respectively.

In the models with $a_{1,0}=0.01$\,AU, as $a_{2,0}/a_{1,0}$ becomes 
greater than $\sim50-60$, there is essentially no decrease in
$t_{\rm merge}$ with respect to the case without the tertiary, 
and all models approach $t_{\rm merge}\sim 5\times10^9$\,yr.
For $a_{2,0}/a_{1,0}\sim20$, there is a minimum in $t_{\rm merge}$.  The retrograde
cases shown with $i=95^\circ$ have very short $t_{\rm merge}$
for small $a_{2,0}/a_{1,0}$, and typically 
merge in a single Kozai timescale $t_{\rm K}$.

Setting the WD radius to be $R_{\rm WD}=5000$\,km, 
for the cases with $a_{2,0}/a_{1,0}=10$ and 20 and $i=89^\circ$, the
periapsis of the inner binary orbit becomes less than the 
$R_{\rm WD}$ in the first Kozai oscillation, at time $t_{\rm K}$.
However, for $R_{\rm WD}=1000$\,km it does not, and the evolution is 
qualitatively similar to that presented in the right
panel of Figure \ref{fig:t}.  Clearly, for these cases a more
complete model with tidal dissipation and circularization is
required to capture the dynamics and to make an accurate 
calculation of $t_{\rm merge}$.  For the purposes of constructing
this figure, I have assumed that $R_{\rm WD}$ is small enough
that a ``collision'' (periapsis $<R_{\rm WD}$) does not occur, 
and thus the results presented may be an upper limit to $t_{\rm merge}$.

The right panel of Figure \ref{fig:wd} shows $t_{\rm merge}$
as a function of $\cos i$ for $a_{1,0}=0.01$\,AU and $a_{2,0}/a_{1,0}=20$, 
30, 40, 50, 60, and 70.  The estimate of equation (\ref{est6}),
which fails to capture the very strong dependence on $a_2/a_1$
is shown as the dashed line.  Again, 
for $a_{2,0}/a_{1,0}\leq20$ and $89\lesssim i_0\lesssim96$,
strong tidal interactions in a single $t_{\rm K}$ are expected;
for these models, as $e_1$ reaches its first maximum, the periapsis 
is less than the fiducial WD radius of $5000$\,km.  However, 
as in the left panel, the results shown assume $R_{\rm WD}$ small
enough that a ``collision'' never occurs.  All other models
have the same qualitative behavior as shown in Figure \ref{fig:t}.

%%%%%%%%%%%%%%%%%%%%%%%%%
\begin{figure*}
\centerline{\includegraphics[width=15cm]{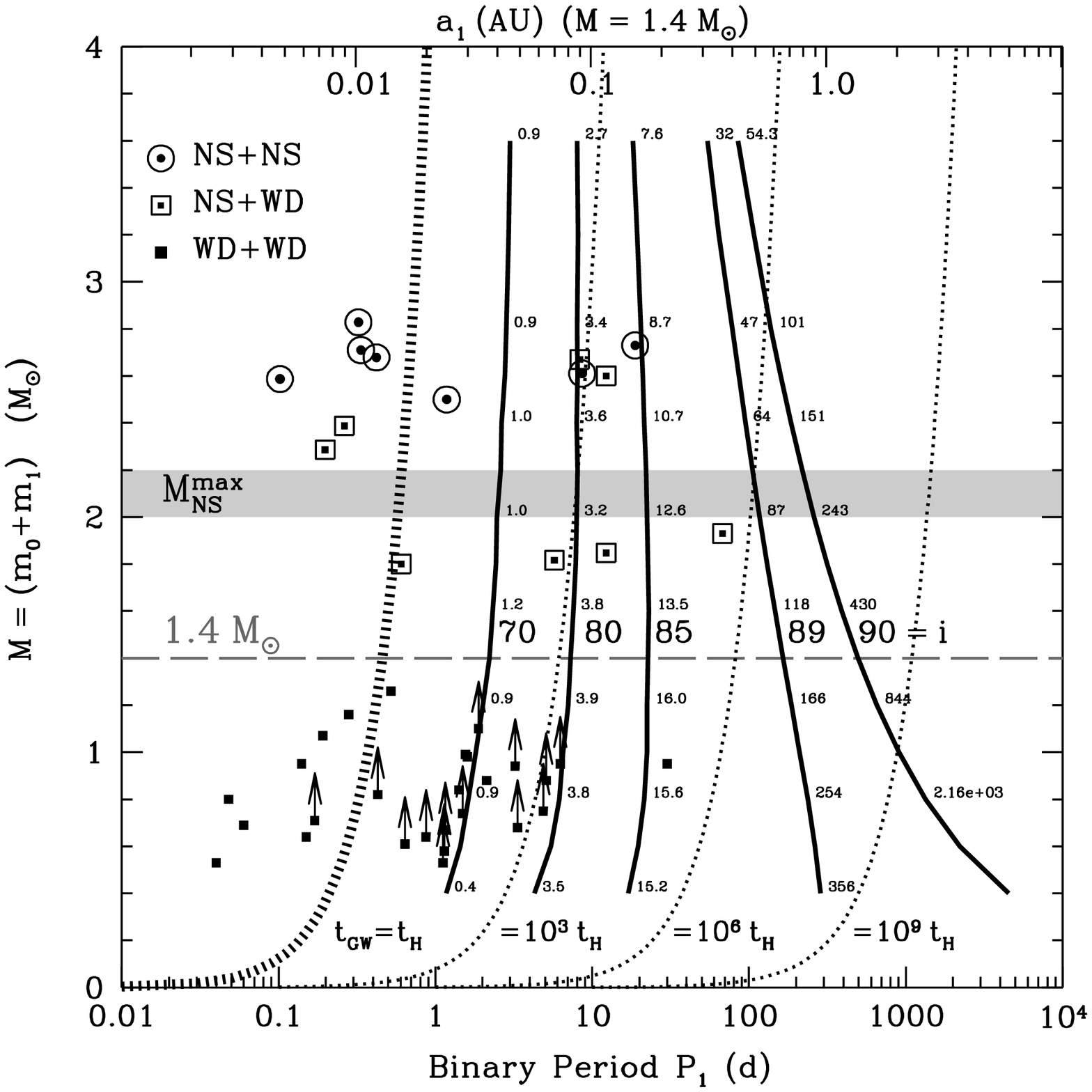}}
\caption{Total mass of compact object binaries $M=m_0+m_1$
versus orbital period (lower axis) and semi-major axis (upper axis,
assuming $M=1.4$\,M$_\odot$).  
The heavy solid lines show the critical value of the 
inner binary period ($P_1^{\rm max}$) such that $t_{\rm merge}=t_{\rm H}$,
including a tertiary with $i_0=70$, 80, 85, 89, and 90 degrees,
assuming $m_0=m_1$, $m_2=1.0$\,M$_\odot$, 
and the same parameters used in Fig.~\ref{fig:ain}.
Each line is labeled with the values 
of the critical tertiary period $P_2$ (yrs) 
for which $P_1^{\rm max}$ occurs (small numbers).
Dotted lines are of 
constant no-tertiary GW merger time of $t_{\rm GW}=1$ (heaviest), $10^3$, $10^6$, and 
$10^9$\,$t_{\rm H}$, assuming  $m_0=m_1$, and $e_1=0$.
Thus, for example, triple systems consisting of binaries with $M=1.4$\,M$_\odot$
(along the horizontal dashed line) 
and $P_1\lesssim500$\,days can in principle merge in $t_{\rm H}$.
For larger $m_2$, the allowed range of $P_1$ increases.  
For some retrograde orbits ($i_0\sim95^\circ$), 
it increases dramatically (see Fig.~\ref{fig:ain}).
Data on WD+WD (filled squares), 
NS+WD (filled+open squares), and NS+NS (filled+open circles) binaries 
are shown (Mullally et al.~2009; Kulkarni \& van Kerkwijk 2010;
Nelemans et al.~2005; Kilic et al.~2010ab, Stairs 2004).  
The Chandrasekhar mass (dashed line) and the 
maximum NS mass (grey shaded) are indicated for reference. 
}
\label{fig:data}
\end{figure*}
%%%%%%%%%%%%%%%%%%%%%%%%% 

%%%%%%%%%%%%%%%%%%%%%%%%%
\begin{figure*}
\centerline{\includegraphics[width=8.5cm]{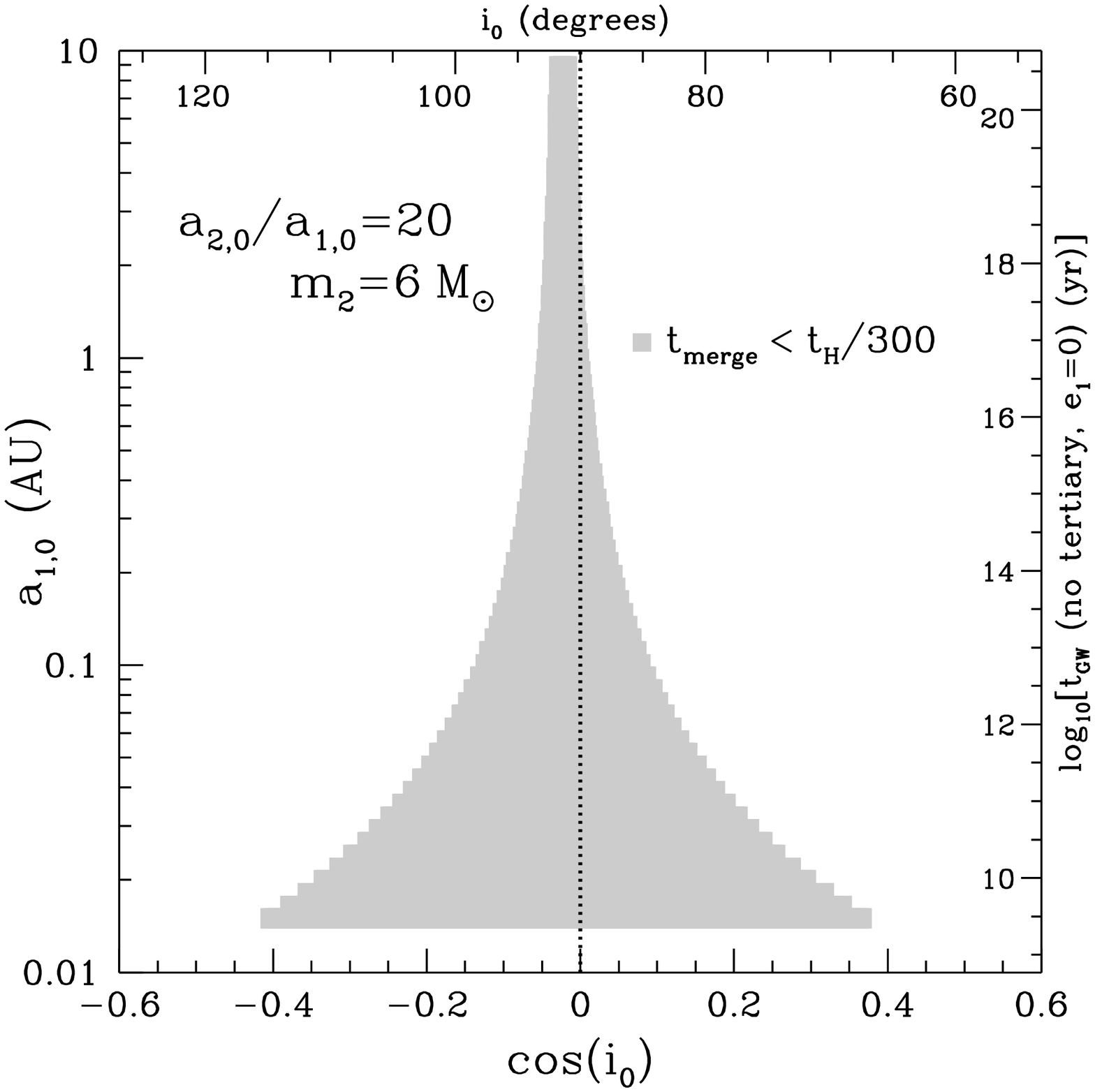}
\includegraphics[width=8.5cm]{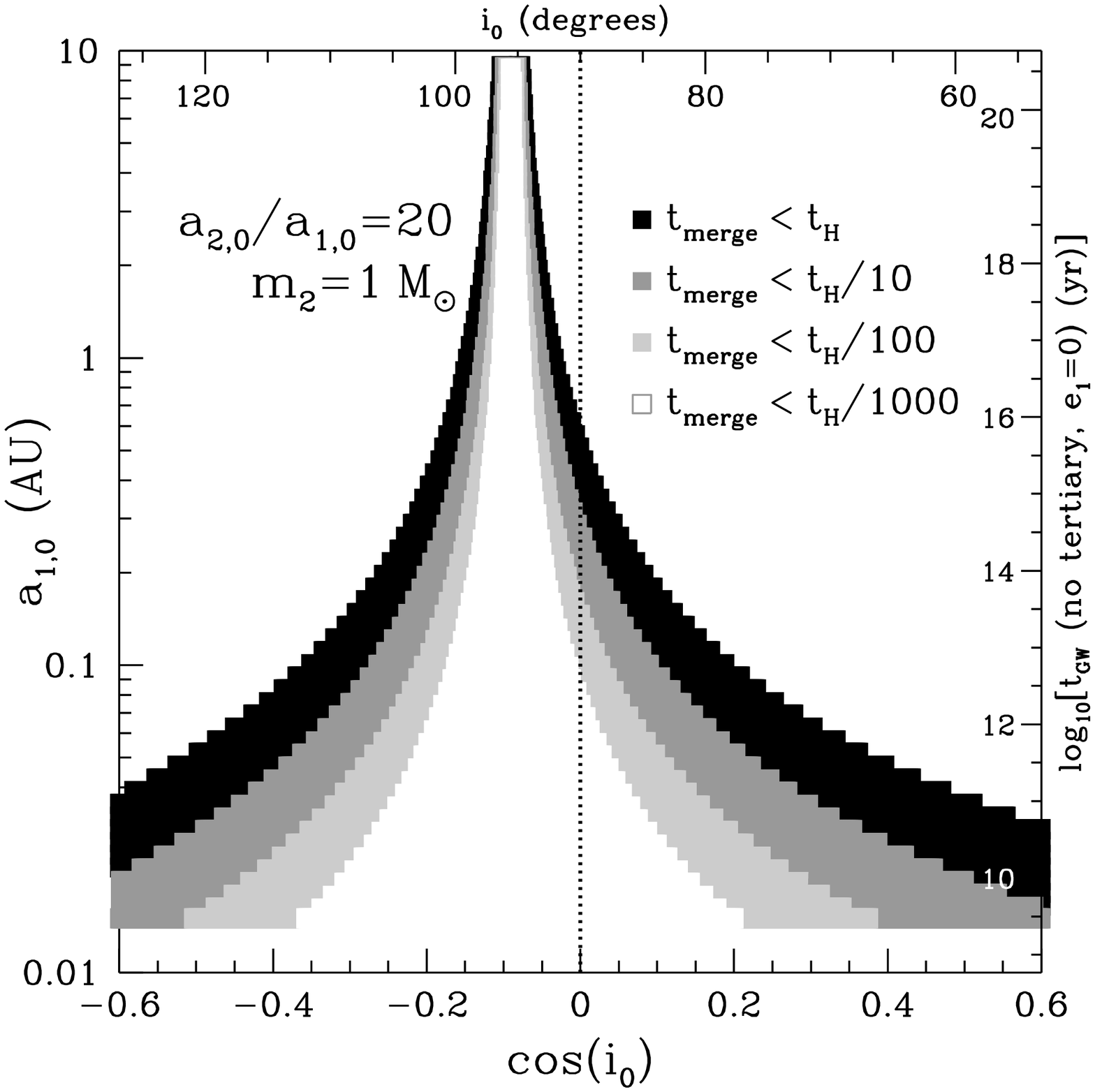}}
\caption{Same as Fig.~\ref{fig:ain},
but for NS-NS binaries with $m_0=m_1=1.4$\,M$_\odot$, $a_{2,0}/a_{1,0} =20$,
and  $m_2=6.0$\,M$_\odot$ (left panel), and $m_2=1.0$\,M$_\odot$ (right panel). 
In the left panel, only $t_{\rm merge}\lesssim t_{\rm H}/300\simeq5\times10^7$\,yr
is shown since this is approximately the main-sequence lifetime of the tertiary.}
\label{fig:ns}
\end{figure*}
%%%%%%%%%%%%%%%%%%%%%%%%%  

In order to make a broad exploration of parameter space
for many models, instead of calculating the detailed time
evolution of each system, as in Figures \ref{fig:t} and \ref{fig:wd},
I use the approximate method described in Wen (2003), which I 
detail in Appendix \ref{appendix:estimate} (see discussion 
after eq.~\ref{est6}).   As shown in the Appendix, the method 
generally underpredicts $t_{\rm merge}$ by a factor of $\sim1.5-2$ 
in most regions of parameter space, but it can overestimate 
$t_{\rm merge}$ by a factor of $\sim10$
for some retrograde cases.  In general, the method is simple, robust,
and accurate over many decades in $t_{\rm merge}$.

Using this approximate method, 
Figure \ref{fig:ain} shows results
analogous to those presented in Figure \ref{fig:wd}, but for 
$m_0=m_1=0.7$, and $m_2=1.0$.  
As a function of the initial value of the inner binary 
semi-major axis, $a_{1,0}$, and as a function of $\cos i_0$, I survey
this parameter space for regions where $t_{\rm merge}$ is 
$<t_{\rm H}$ (black), $<0.1t_{\rm H}$ (dark grey), $<0.01t_{\rm H}$
(light grey), and  $<0.001t_{\rm H}$ (interior white) for initial values of
$a_{2,0}/a_{1,0}=20$ (upper left), 40 (upper right), 60 (lower left), and 80 (lower right).  
Only regimes where the nominal
GW merger timescale without the tertiary is larger than $t_{\rm H}$
are explored ($a_{1,0}\gtrsim0.015$\,AU; see eq.~\ref{tgw}).  The
right vertical axis gives the associated no-tertiary GW merger time; 
considered values for $a_{1,0}$ run up to 10\,AU, equivalent to 
$t_{\rm GW}\sim10^{22}$\,yr without the tertiary.
The basic trends presented in the 
dynamical calculations shown in Figure \ref{fig:wd} are reproduced.
Exceedingly rapid mergers can be induced for a narrow range of retrograde 
values of $i$ near 95$^\circ$, and for many prograde tertiary orbits.  
As $a_{2,0}/a_{1,0}$ increases, the allowed
area in the $a_{1,0}-\cos i_0$ plane decreases, and for 
$a_{2,0}/a_{1,0}\gtrsim 100$, Kozai oscillations do not accelerate mergers; 
this follows from the strong dependence
of $t_{\rm K}$ on $a_2/a_1$ and the effects of GR precession.  
Nevertheless, it is clear
from Figure \ref{fig:ain} that for a significant region of parameter space,
$t_{\rm merge}$ can be less than 10, 1, 0.1, or even 0.01\,Gyr, 
even for systems that have nominal (no tertiary) 
merger times of $\gg t_{\rm H}$.

Using these estimates of $t_{\rm merge}$, it is possible to 
calculate the maximum possible value of the inner binary semi-major axis 
$a_{1,0}^{\rm max}$  such that merger occurs in a single Hubble time,
given a tertiary of mass $m_2=1.0$\,M$_\odot$ at any $i$.  
Figure \ref{fig:ain} implies that for retrograde orbits $a_{1,0}^{\rm max}$
is extremely large near $i_0\sim95^\circ$.  If we restrict our attention
to prograde orbits, there is a unique value of $a_{1,0}^{\rm max}$ for each
initial $i_0$ such that $t_{\rm merge}\leq t_{\rm H}$.  As implied by
the left panel of Figure \ref{fig:wd}, since $t_{\rm merge}$
exhibits a minimum as a function of $a_{2,0}/a_{1,0}$, $a_{1,0}^{\rm max}$,
will occur not at the smallest $a_{2,0}/a_{1,0}$, but instead at 
an intermediate value determined by the shape of the 
minimum in $t_{\rm merge}(a_{2,0}/a_{1,0})$.  

In Figure \ref{fig:data}, the solid lines show the critical maximum 
inner binary period $P_1^{\rm max}$ 
corresponding to this $a_1^{\rm max}$ as a function of
$M=m_0+m_1$, for tertiary mass of $m_2=1.0$\,M$_\odot$ and 
inclination of 70$^\circ$, 80$^\circ$, 85$^\circ$,  89$^\circ$, and 90$^\circ$.   
Period is shown instead of semi-major axis
to make contact with observational plots of compact binaries
prevalent in the literature.  The top axis gives $a_1$ for 
$M=1.4$\,M$_\odot$ for reference. The dotted lines 
are of constant GW merger timescale assuming $e_1=0$
and $m_0=m_1$ for multiples of $t_{\rm H}$.  
In the absence of a tertiary, 
for $M=1.4$\,M$_\odot$ (dashed line)
a binary period of $\lesssim0.4$\,days is required for 
merger in $t_{\rm H}$.  The solid lines show that 
the allowed range in observed inner binary period 
is dramatically increased for binaries with a
hierarchical tertiary.  The small numbers along the 
solid lines denote the value of the tertiary period
$P_2$ (using $M+m_2$) at each value of  $P_1^{\rm max}$ ($a_1^{\rm max}$) for each $i$.
Again taking a Chandrasekhar mass inner binary with $M=1.4$\,M$_\odot$
(following the dashed line),
for $i=70^\circ$,  I find that $a_1^{\rm max}\simeq0.037$\,AU, 
$P_1^{\rm max}\simeq2.2$\,days, and $a_2/a_1\simeq37$,
implying that the 1\,M$_\odot$ tertiary is required to 
be at $a_2\simeq1.37$\,AU, with an orbital period of 
$P_2\simeq380$\,days (using $M+m_2=2.4$\,M$_\odot$). As another example,
for  $i=85^\circ$,
$a_1^{\rm max}\simeq0.175$\,AU, $P_1^{\rm max}\simeq22.7$\,days,
$a_2/a_1\simeq46$, $a_2\simeq8$\,AU, and 
$P_2\simeq5400\,{\rm days}\,\simeq15$\,years.

Figure \ref{fig:data} implies that for any given observed WD-WD
binary with $M\simeq1.4$\,M$_\odot$, in order to exclude the
possibility that the system will merge within one Hubble time,
one must exclude the possibility of a tertiary with some 
mass $m_2$, at some inclination $i$.  Given the very large 
values of $P_2$ for $i\gtrsim80^\circ$, this may be observationally
challenging. 
Almost all of the WD-WD binaries could in
principle merge in less than $t_{\rm H}$ if one posits the
existence of a $\sim1$\,M$_\odot$ tertiary at $\gtrsim 80^\circ$ and 
with $a_2/a_1\simeq40$ ($P_2\sim4$\,yr).
Note that the values for $P_1^{\rm max}$ shown in Figure \ref{fig:data} 
shift to yet longer periods if $m_2$ increases.  For retrograde orbits, 
particularly near $\sim95^\circ$ (see Fig.~\ref{fig:ain}), they increase 
still further.  Again, some caution is warranted in interpreting
the solid lines in Figure \ref{fig:data}  
for $i\gtrsim89^\circ$ since for WD-WD binaries tidal
interactions will almost certainly affect the estimate of $t_{\rm merge}$.
Nevertheless, it is difficult to see how tides would dramatically
increase  $t_{\rm merge}$ when the relevant comparison is to $t_{\rm H}$.

\subsection{NS-NS Binaries}

Similar plots to Figure \ref{fig:ain} may be generated for the 
case of NS-NS mergers.  Assuming $m_0=m_1=1.4$\,M$_\odot$, 
the left panel of Figure \ref{fig:ns}
shows the region in the $a_{1,0}-\cos i_0$ plane for which a 
merger occurs in $\sim5\times10^7$\,yr $\simeq t_{\rm H}/300$,
which approximates the main-sequence lifetime of the tertiary,
assumed here to be $m_2=6$\,M$_\odot$.  The right panel shows
the same quantities, but for $m_2=1$\,M$_\odot$, as might be 
appropriate after the tertiary becomes a WD.
As discussed in \S\ref{section:ns_scenario}, in some 
cases the tertiary might also be a massive BH, in which 
case the allowed region for which $t_{\rm merge}$ is 
less than $t_{\rm H}$ grows.
Similar to the WD-WD case, Figure \ref{fig:data} shows that 
for a 1\,M$_\odot$ tertiary with $i\gtrsim85^\circ$ and
$P_2\lesssim10$\,yrs, all the observed NS-NS systems
merge in $t_{\rm H}$ given an appropriately placed tertiary.

The parameter space for which the tertiary remains bound to 
the system through both of the supernovae that produce the 
NS-NS binary may not be large.  In particular, the tertiary
can neither be too close, because of dynamical stability, nor too 
far way because of the binding energy of the orbit relative to 
the kick imparted to the center-of-mass during each of the
inner binary component's supernovae.
However, inspection of the plots by Kalogera (1996)
for the binary case indicate that
for relatively small tertiary semi-major axis ($a_2\lesssim$\,few AU),
an intermediate-mass tertiary can stay bound, and perhaps 
acquire high eccentricity during the supernovae for kick velocities of
$\sim100$\,km s$^{-1}$ in some cases.  A high value of $e_2$
could lead to more rapid mergers than implied by Figure
\ref{fig:ns}. As discussed
in Sections \ref{section:wdwd} and \ref{section:ns_scenario},
$a_2\sim$\,AU likely guarantees strong (common envelope) 
interaction between 
all components as the primary and then the secondary evolve 
off the main sequence.

\section{Discussion}
\label{section:discussion}

\subsection{Rates}

The fact that essentially all close solar-type binaries are
in triple systems argues that their subsequent
WD-WD binaries will also be in triple systems. 
Depending on the semi-major axis and inclination
distribution of the tertiaries at the time of formation of the
compact objects
one expects Kozai oscillations to speed up the process of coalescence
of the inner binary significantly,
as shown in Figures \ref{fig:t}-\ref{fig:ain}. 
 The work of FT07, together with Figure \ref{fig:ain},
serve as a guide to an estimation of the rate, but the estimate is 
complicated by the expectation that the distribution of 
inclinations may  be biased towards co-planar orbits for coeval systems.  

Nevertheless, to make a simple
rate estimate, assume that the distribution of
binary and tertiary semi-major axes is flat (equal numbers in 
log semi-major axis), and that the probability of having a 
system at inclination $i$ is $dp/di=\sin i$.  Importantly,
Figure \ref{fig:ain} shows that the range of $a_1$ strongly affected
by Kozai oscillations is comparable to the relevant range of {\it close} binary 
semi-major axes.  In addition, the Kozai
mechanism operates over $\sim$\,one decade in $a_2/a_1\sim 3-100$,
which corresponds to $\sim2$ decades in $a_2$. Thus, one then expects
many of the triples to be strongly affected.  
Finally, although the area in the $a_1-\cos i$ plane is smaller for
more rapid mergers, the systems at large $i$ dominate the rate.
For example, comparing the prograde regions of the upper left panel
of Figure \ref{fig:ain}, one sees that the total area of the shaded
regions (with $a_1$ measured in log units) measures the fraction of 
all systems with that merger time.  Although the area of 
the black region is roughly 10 times that of the white, the latter 
merge 1000 times faster.  Thus, the average rate is dominated 
by the highest inclination systems, a consequence in part of the 
very steep scaling of $t_{\rm merge}$ with $\cos i$ (e.g., eq.~\ref{est6},
Fig.~\ref{fig:wd}).  This can be shown explicitly by using the 
very crude estimate of equation (\ref{est6}).  Momentarily ignoring
the dependence on $a_2$, one may write the total merger rate as
\beq
\frac{dN}{dt}\sim\frac{153}{100}
\left(\frac{Mm_0m_1G^3}{a_1^4 c^5 \,\cos^6 i}\right)
\left[\frac{dN}{d\ln a_1}+\frac{2}{3}\frac{dN}{d\ln \cos i}\right],
\label{rate1}
\eeq
where $N$ is the total number of systems.
As in equation (\ref{est6}), this expression neglects the 
very important $m_2$ dependence of the Kozai mechanism and 
the $a_2/a_1$ dependence, and thus equation (\ref{rate1})
should be considered schematic.  For practical purposes
the dependence on $a_2/a_1$ can be considered 
nearly a step function since for all relevant parameters
Kozai is ineffective for $a_2/a_1\gtrsim100$, as shown in 
Figures \ref{fig:wd} and \ref{fig:ain}. As discussed in 
Section \ref{section:scenarios}, the latter is particularly
important since the inner binary will likely undergo
two common envelope events and significant mass loss, thus 
increasing $a_2/a_1$ from its initial value. Nevertheless,
if triple systems are prevalent, one expects them to contribute 
both to the very prompt rate ($<10^8$\,yr; \S\ref{section:prompt}), and 
to the delayed rate ($>$\,Gyr), since for $i\lesssim39.2^\circ$ (and/or
$a_2/a_1>100$ and/or $m_2\ll1$\,M$_\odot$), Kozai
does not affect the merger time of the inner binary. 
Thus, for low-$i$ or large $a_2/a_1$ systems, 
one should obtain a delay-time distribution
identical to the binary-only case.  A more
complete calculation of the delay-time distribution of triple
systems is saved for a future work.

\subsection{How ``prompt'' is prompt?}
\label{section:prompt}

A prompt component to the supernova Ia rate has been claimed
by a number of authors (e.g., Scannapieco \& Bildsten 2005;
Mannucci et al.~2006; Aubourg et al.~2008; 
Brandt et al.~2010; Maoz \& Badenes 2010; Maoz et al.~2010; Maoz et al.~2011).  
Various estimates suggest that $\sim50$\% of Ia's are ``prompt,''
with a characteristic short timescale of $\sim0.1-1$\,Gyr,
and potentially with a two component power-law or bimodal delay-time distribution
(see Mannucci et al.~2006 and the recent review by Maoz 2010).  

In the picture presented here, a fraction of intermediate-mass
stars are born in triple systems.  The fraction with high
inclination tertiaries can merge extremely rapidly as soon
as both stars in the inner binary are WDs.  Because the 
calculated merger times are in many cases much less than
even the post-main-sequence timescales of these stars,
one expects the fastest WD-WD mergers to be
limited only by the main-sequence lifetime of their progenitors. 
Thus, $\sim8+8$\,M$_\odot$ binaries in triple systems merge first,
immediately after WD birth, and thus the minimum delay time 
is $\sim3\times10^7$\,yr.  In this case, one expects a delay-time
distribution of $\sim t^{-1/2}$  (Pritchet et al.~2008). 
These prompt supernovae would preferentially be super-Chandrasekhar 
mass binaries (see Pakmor et al.~2010). 

Many of these statements are equally applicable to NS-NS
mergers.  For example, Figure \ref{fig:ns}
shows that the merger timescale for NS-NS binaries in 
triple systems can be very short.  Again, there is a 
region of parameter space where the limiting factor
is the time required to produce the NS-NS binary,
again implying that the fastest NS-NS
mergers can come just $\sim10^7$\,yr
after the last star formation episode.  This
may explain the fact that many short-duration GRBs
are seen in star-forming galaxies (e.g., Berger 2009).
By extension, one expects similar arguments to hold
for BH-WD/NS or NS-WD systems that in some cases might 
give rise to long-duration GRBs (Fryer et al.~1999).

These considerations are amplified if the 
multiplicity of stars increases with zero-age main 
sequence mass (Lada 2006; Raghavan et al.~2010).

\subsection{Progenitors}

Even if they constitute much of overall merger rate, 
the progenitor systems for the shortest-lived
compact object mergers will be very rare in a given 
galaxy simply because the time a given system spends 
as a progenitor is very short.   Nevertheless, as in 
single-degenerate models of Ia supernovae, for WD-WD
systems with Gyr merger time (with the tertiary!), 
one could look for close binaries 
with evidence of a tertiary companion.  This would
work if the tertiary is a WD or NS, but the problem
is that in many cases one expects the companion to 
be a main-sequence star of $\sim1$\,M$_\odot$ (\S\ref{section:scenarios}).
The latter would necessitate a new search strategy,
since most searches for close WD-WD binaries are color-selected
(Napiwotzki et al.~2001; Badenes et al.~2009; Brown et al.~2010).
One example would be to target intermediate mass main sequence
stars for radial velocity measurements and search for the signal of a 
more massive, but unseen companion, which might be an old, high-mass 
WD-WD (or NS/BH) binary.

The same is true of the coeval NS-NS systems
described in \S\ref{section:scenarios}.  In that case,
one might look for NS-NS binaries in pulsar searches,
with timing characteristics that suggest the presence of a
tertiary component.  However, in many cases the NS
components may not be observed as pulsars, and the tertiary may
be a main sequence $\sim6-8$\,M$_\odot$ star.
In both the WD-WD and NS-NS cases, the only visible
progenitor at the sight of the subsequent explosion 
(the compact object merger) in pre-explosion imaging
may be a bright main sequence star (similar in spirit and 
conclusion to the recent work by Kochanek 2009).

\subsection{Transients}

Recently, several authors have discussed the possibility 
that WD-WD collisions, at small impact parameter, may be a
way of producing Ia-like supernovae (e.g., Rosswog et al.~2009;
Raskin et al.~2010).  For some of the calculations presented
in Figures \ref{fig:wd} and \ref{fig:ain} the merger occurs
in a single Kozai timescale, at very high eccentricity.
I have neglected tidal forces on the compact objects 
throughout this work (\S\ref{section:method}), but these
findings suggest that in some cases something akin to a
``collision,'' or at least a very strong tidal interaction,
may be induced by the Kozai mechanism.  This is 
particularly promising for retrograde orbits with 
$i\sim95^\circ$ (Fig.~\ref{fig:ain}).  Such tertiaries
may be captured in binary-binary collisions in dense
stellar environments (Ivanova et al.~2008, 2010),
or perhaps in some cases may be coeval (see Fig.~7 of FT07).
Simple numerical experiments with the evolutionary equations
described in \S\ref{section:method}, but including
apsidal motion of the inner binary as a result of 
tides (eq.~\ref{ttide}), still allow for a region of parameter space
where the merger occurs at very high eccentricity.

%%%%%%%%%%%%%%%%%%%%%%%%%
\begin{figure*}[h]
\centerline{\includegraphics[width=8.4cm]{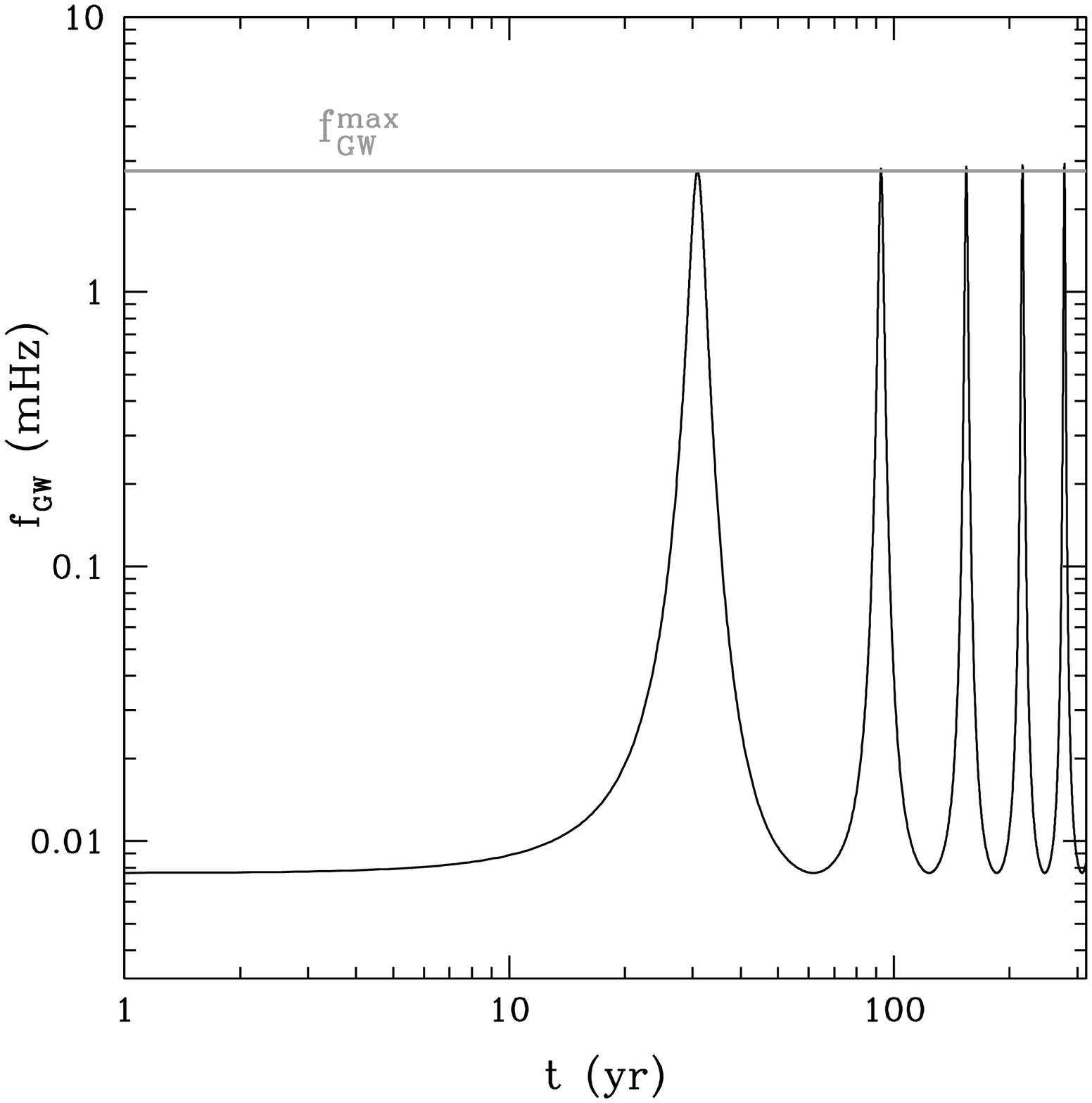}
\includegraphics[width=8.4cm]{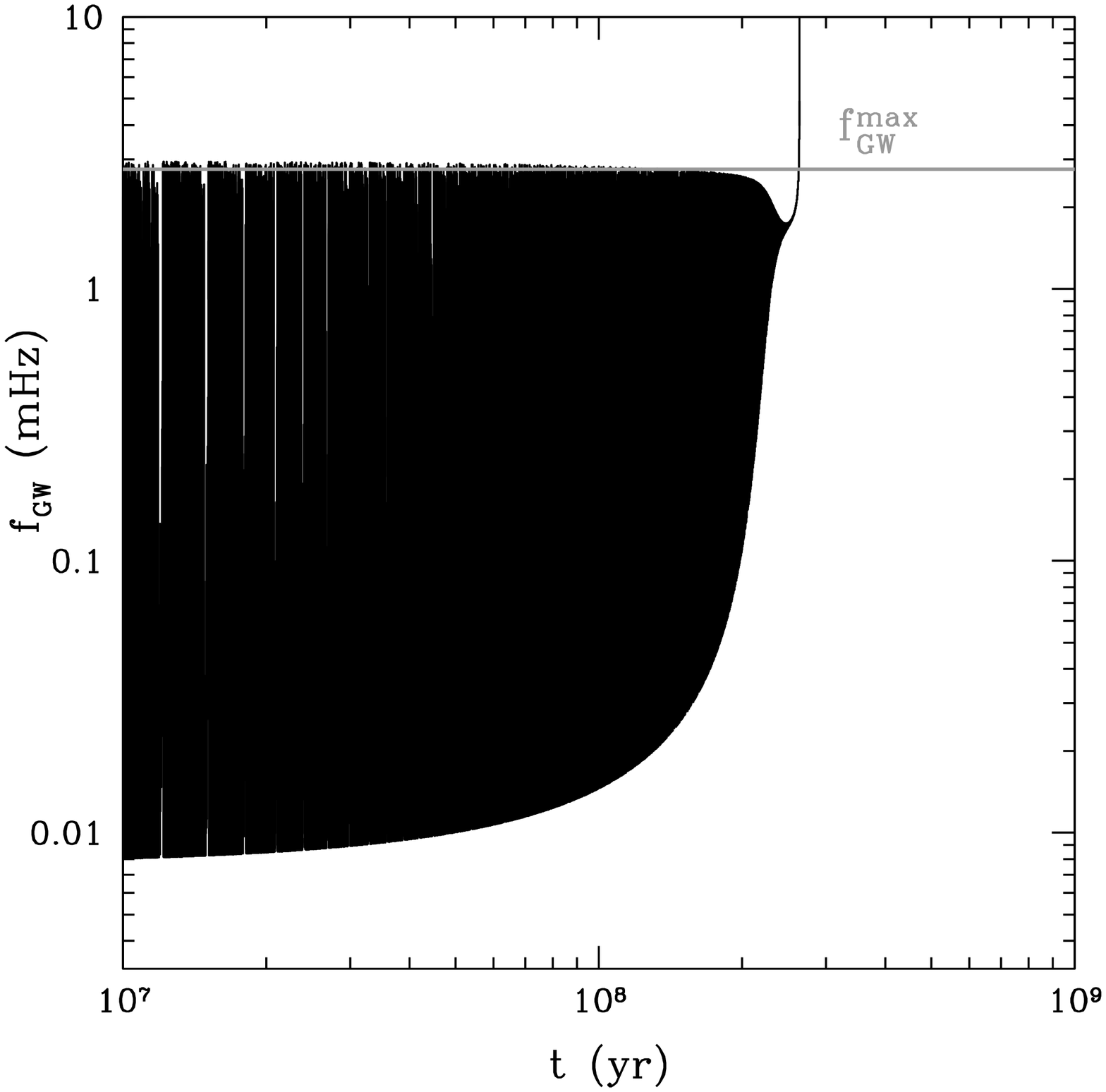}}
\caption{Time evolution of the peak GW frequency 
$f_{\rm GW}(m{\rm Hz})$ (eq.\ \ref{fgw}) for the system shown in Figure \ref{fig:t}.
The gray line shows the maximum GW frequency,
obtained by combining equation (\ref{fgw}) with the
estimate of the maximum eccentricity given in Appendix \ref{appendix:estimate}. }
\label{fig:gw}
\end{figure*}
%%%%%%%%%%%%%%%%%%%%%%%%% 

%%%%%%%%%%%%%%%%%%%%%%%%%
\begin{figure*}[h]
\centerline{\includegraphics[width=8.4cm]{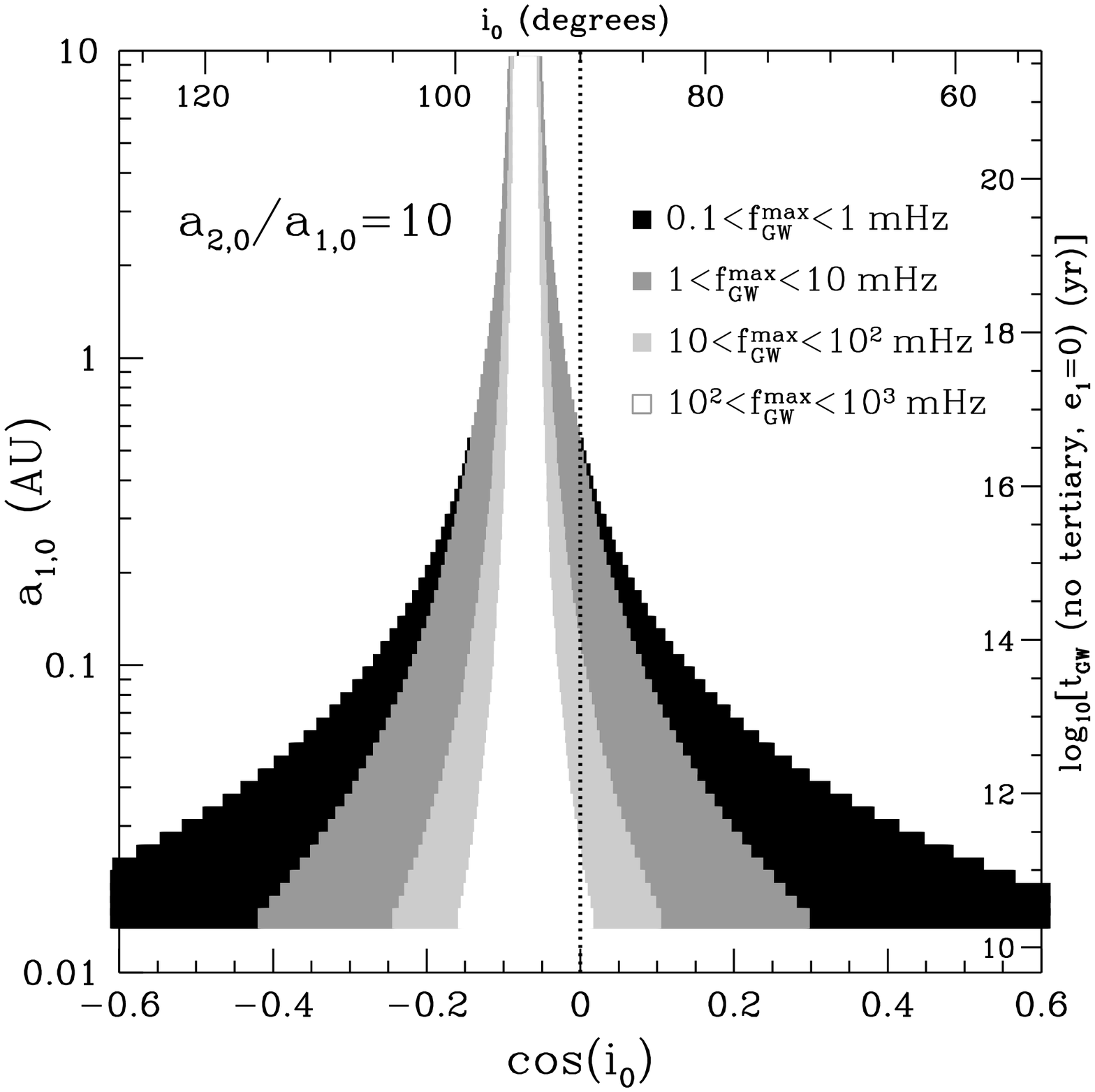}
{\includegraphics[width=8.4cm]{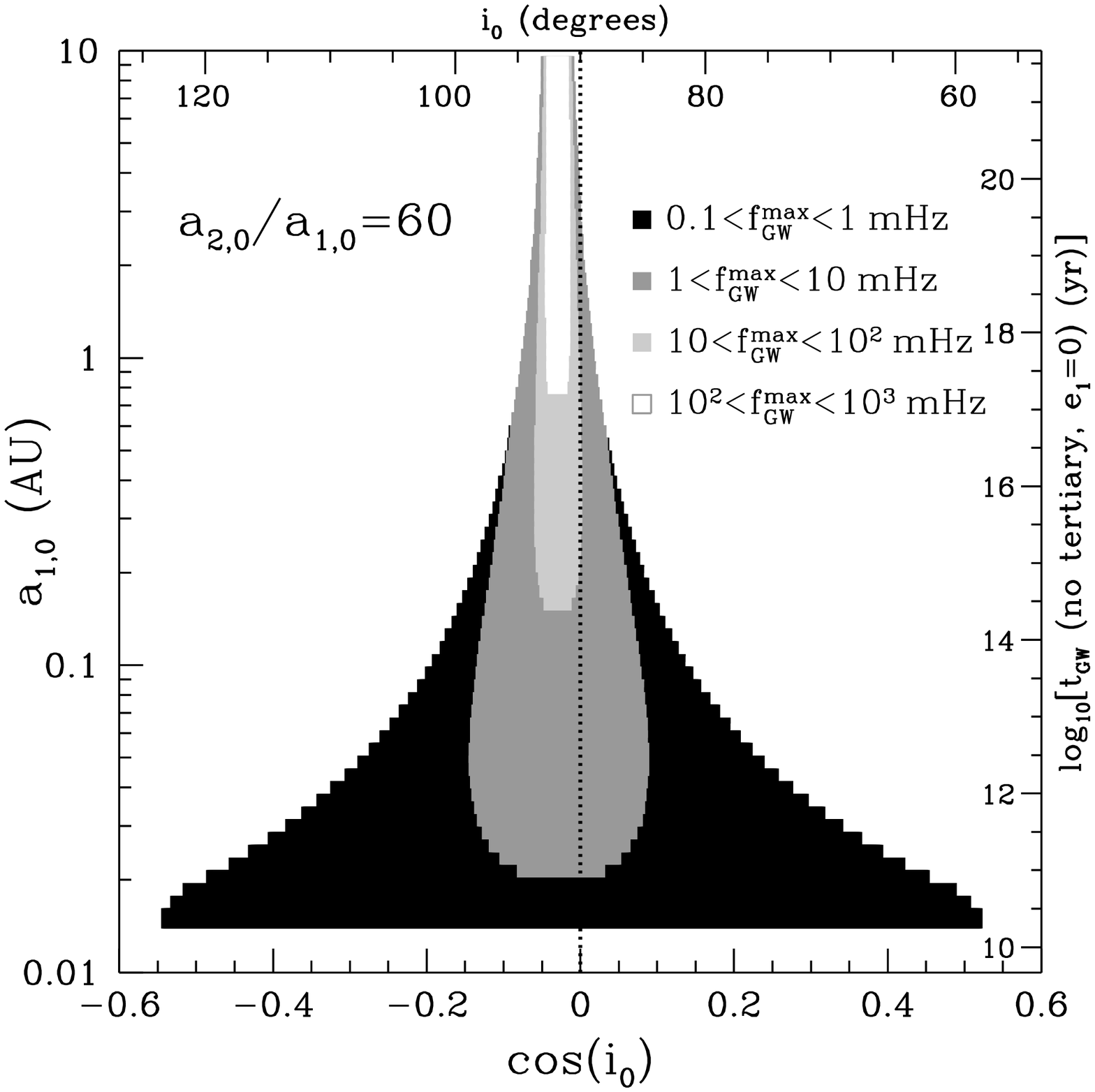}}}
\caption{The same as the top and bottom left panels of Figure \ref{fig:ain},
respectively, but showing regimes for which the maximum value of the 
GW frequency $f_{\rm GW}^{\rm max}$ (as computed from
eq.~\ref{fgw} and Appendix \ref{appendix:estimate}) 
is in the range $0.1-1$ (black), $1-10$ (dark gray),
$10-10^2$ (light gray), and $10^2-10^3$\,mHz (white).
The latter will be affected by tidal interaction
between the WDs, which is not captured by the calculations presented. }
\label{fig:f}
\end{figure*}
%%%%%%%%%%%%%%%%%%%%%%%%% 

\subsection{Interactions \& The Tertiary}

{\it WD-WD mergers:}  If WD-WD mergers driven by a hierarchical 
tertiary produce Type Ia supernovae, the explosion should 
interact with and overtake the tertiary.  Thus, just as in 
searches for the remaining star in the single-degenerate
scenario (e.g., Kerzendorf et al.~2009), 
in the triple scenario proposed here, the tertiary
in historical Galactic supernovae should be currently inside
the Ia supernova remnant.  In addition, depending on the structure
of the tertiary, some amount of mass might be expected to be 
lost by the shockwave interaction (Marietta et al.~2000).  However,
in contrast to the single-degenerate picture (Whelan \& Iben 1973), 
one expects the tertiary to be in many cases more than an AU 
distant from the explosion and would only very rarely be a 
giant. Using the simple analytic 
model of mass stripping and ablation in supernova explosions
with binary companions developed by Wheeler et al.~(1975),
I find that even in the most compact cases considered here,
with $a_2\sim0.1$\,AU, the total mass stripped and ablated 
from a main sequence solar mass tertiary is just $\sim0.01$\,M$_\odot$.
 For $a_2\gtrsim1$\,AU, the expected
mass lost from the tertiary will be minimal.\footnote{The
fraction of the mass stripped from the tertiary is 
$\sim10^{-3}(M_{\rm SN}/{\rm M}_\odot)({\rm M}_\odot/m_2)[(R/{\rm R}_\odot)/(a_2/0.1{\rm AU})]^2$
$(v_{\rm SN}/10^4\,{\rm km/s})(600\,{\rm km/s}/v_{\rm esc})$, where $M_{\rm SN}$
is the mass of the supernova ejecta, $v_{\rm SN}$ is the ejecta velocity, and 
$v_{\rm esc}$ is the escape velocity from the tertiary.
The mass shock heated and ablated from the tertiary is $\sim6$ times higher
(see Tab.~1 of Wheeler et al.~1975).}  Similarly, based on the work
of Kasen (2010), one expects a soft X-ray flash as the Ia shockwave
overtakes the tertiary on a timescale of $\sim10^4$\,s\,\,$(a_2/{\rm AU})$
for a shockwave of $\sim10^4$\,km s$^{-1}$ and a very small change to the 
early-time optical and UV lightcurve.

{\it NS-NS mergers:}
If NS-NS mergers in triple systems 
produce short-duration gamma-ray bursts (SGRBs), then there is a
chance for interaction of the relativistic blastwave with 
the tertiary companion because the systems most likely
to be affected by Kozai oscillations are at high mutual
inclination relative to the inner binary, and because the
inferred opening angle for SGRBs is relatively large 
(e.g., $10-30^\circ$; Nakar 2007). 
One then envisions an interaction
similar to that calculated in MacFadyen et al.~(2005),
but with either a main-sequence, WD, or BH companion
to the SGRB explosion
(see \S\ref{section:ns_scenario}, Fig.~\ref{fig:ns}).
The timescale for interaction is $\sim a_2/c\sim500(a_2/{\rm AU})$\,s.
Moreover, because we are presumably looking roughly down the jet
axis, the tertiary may in some cases be roughly 
along the line of sight. In the case of a main-sequence
companion, some material may be stripped and ablated.  However,
as in the WD-WD case, $a_2$ is expected to be in the range 
$\sim0.1-10$\,AU, and this may preclude a large effect on the 
tertiary, or the SGRB lightcurve.

\subsection{Gravity Waves}

{\it The Diffuse Background:}
If most close WD-WD binaries were born in triple systems,
predictions for the gravitational wave background may be 
modified from fiducial predictions (e.g., Farmer \& Phinney 2003).
For highly eccentric orbits
the total power in GWs, and the peak frequency of GWs, increases 
significantly.  The peak GW frequency is well approximated by
(Wen 2003)
\beq
f_{\rm GW}^{\rm max}
=\frac{1}{\pi}\left(\frac{GM}{{a^3_1(1-e_1^2)^{3}}}\right)^{1/2}(1+e_1)^{1.1954}.
\label{fgw}
\eeq
As an example, 
Figure \ref{fig:gw} shows the early (left panel) and late (right panel)
time evolution of the GW frequency for the triple system 
shown in Figure \ref{fig:t}.    Although the inner binary,
has $f_{\rm GW}\simeq0.008$\,mHz in the absence of a tertiary,
with the addition of the tertiary $f_{\rm GW}$ increases strongly,
and periodically, on a timescale $t_{\rm K}$ (eq.~\ref{tk}).

As shown in Appendix \ref{appendix:estimate},
the maximum eccentricity attained by the inner binary
can be estimated semi-analytically.  It is these estimates
that are used in the calculations of the merger time shown in 
Figures \ref{fig:ain} and \ref{fig:ns}.  Substituting
this maximum eccentricity into equation (\ref{fgw})
yields $f_{\rm GW}^{\rm max}$, the maximum GW frequency 
produced by the binary during inspiral.  The horizontal
solid line in Figure \ref{fig:gw} shows this estimate for 
$f_{\rm GW}^{\rm max}$.

Figure \ref{fig:f} shows a summary of results for 
$f_{\rm GW}^{\rm max}$ computed from the results for 
two panels of Figure \ref{fig:ain} 
for WD-WD binaries ($a_{2,0}/a_{1,0}=10$, 60).  
For the models that merge in less than $t_{\rm H}$, 
regions of different peak $f_{\rm GW}^{\rm max}$
are shaded in the $a_{1,0}-\cos i_0$ plane.  All 
frequencies shaded are in the range detectable by {\it LISA}.
Note that the WD-WD binaries that attain the highest frequencies
(interior white) have the highest peak eccentricities,
the smallest periapses, and will interact
tidally, thus modifying the results presented here (\S\ref{section:numerics}).
This can be seen since $f_{\rm GW}^{\rm max}$
approaches the inverse dynamical time for an individual 
WD for some regions of parameter space.

Since the extragalactic GW background is dominated by close WD-WD binaries
(Farmer \& Phinney 2003),
which are generally assumed to be circular as a result of the
preceding common envelope evolution, 
if the fraction of  close binaries that are actually in triple systems is 
fairly large, as implied by the statistics on solar-type
binaries (Tokovinin et al.~2006; FT07), then the expected GW
background will be modified. 
Recent studies of the {\it LISA} GW foreground focus exclusively on 
circular WD-WD binaries (Ruiter et al.~2010), and these results
too would need to be revisited if the triple fraction of 
close WD-WD binaries is high.  The only cases where eccentric
WD-WD binaries have been considered are in globular clusters
where high eccentricity can be imparted to binaries during
binary-single and binary-binary interactions (Benacquista 2001; Ivanova et al.~2006;
Willems et al.~2007).

{\it Sources \& Foreground:}
Whether NS-NS, WD-WD, or otherwise, 
the most eccentric systems will be short-lived, and this decreases 
the probability that they can be seen.  The competing effect,
of course, is that their overall GW luminosity is larger.
The right panel of Figure \ref{fig:gw} indicates that a
$a_1=0.05$\,AU WD-WD binary in a triple system with a
1\,M$_\odot$ tertiary at $a_2/a_1=20$ would go through periodic 
decade-long enhancements in $e_1$, that are potentially
detectable by {\rm LISA} for nearby systems.  During the mission itself, 
many such binaries might be seen, both Galactic WD-WD systems with 
larger $a_1$ and NS-NS systems with higher $f_{\rm GW}^{\rm max}$.
The effect of the Kozai mechanism is to make otherwise 
unobservable Galactic WD-WD binaries observable during pericenter 
passage, since it is at these times at high $e_1$ that 
$f_{\rm GW}^{\rm max}$ is maximized.  This may cause
individual sources to be observable at pericenter passage,
and it may cause an added degree of complexity to the 
GW foreground from all local sources.  See Gould (2011).

\acknowledgments 

I am grateful to Omer Blaes for sharing the code from Blaes et al.\ (2002)
for the purposes of initially validating the results of the code presented 
here.  In addition, I thank Andy Gould, Chris Kochanek, and Kris Stanek for 
discussions and encouragement. Additional discussions with Ond\v{r}ej Pejcha,
Benjamin Shappee, and Brian C.~Lacki are acknowledged.  This work is supported in part by an Alfred 
P.~Sloan Foundation Fellowship and NSF grant AST-0908816.

%------------------------------------------------------------------------------

\begin{appendix}

\section{Approximation to the Merger Time}
\label{appendix:estimate}

I use the methods discussed in MH02 and Wen (2003) to make
simple, but accurate, estimates of $t_{\rm merge}$ to supplement
the direct calculation of the time-dependence of the orbital
elements, as described in \S\ref{section:method}.  Since this 
scheme provides an efficient way to estimate the merger timescale
for a very wide range of system parameters, as in Figures
\ref{fig:ain} \& \ref{fig:data},
I repeat the steps here.

The goal is to estimate $t_{\rm merge}$ from the initial conditions
of the triple system.  Following Wen (2003), I use the fact that 
(neglecting gravitational radiation), the
quadrupole-level Hamiltonian is conserved throughout the evolution.
It can be written in terms of $\epsilon=1-e_1^2$, $i$,  and 
$g_1$ as (MH02)
\begin{equation}
W(\epsilon,g_1)=-2\epsilon+\epsilon
\cos^2i+5(1-\epsilon)\sin^2g_1(\cos^2i-1)+
\frac{4}{\sqrt{\epsilon}}\frac{(m_0+m_1)}{m_2}
\left(\frac{b_2}{a_1}\right)^3
\left(\frac{2G(m_0+m_1)}{a_1c^2}\right),
\label{ham}
\end{equation}
where $b_2=a_2(1-e_2^2)^{1/2}$, and the
last term is the 1st-order post-Newtonian correction that 
accounts for GR precession.

Starting with initial values of the eccentricity of the 
inner binary and its argument of periastron, $e_{1,0}$ ($\epsilon_0$)
and  $g_{1,0}$, as well as the initial inclination $i_0$,
the system evolves to a maximum eccentricity
$e_{1,\rm \, max}$ and thus minimum $\epsilon_{\rm min}$,
at a critical $g_{1,\rm\,crit}$ and $i_{\rm crit}$.  
Taking $d\epsilon/dt=0$ at the moment $\epsilon=\epsilon_{\rm min}$
provides a relationship between $\epsilon_{\rm min}$, $g_{1,\rm\,crit}$, 
and $i_{\rm crit}$ (Wen 2003):
\beq
\sin(2g_{1,\,\rm crit})=
\frac{8}{225}\left(\frac{G(m_0+m_1)}{a_1c^2}\right)^{3/2}
\left(\frac{Gm_0m_1}{a_1c^2m_2}\right)\left(\frac{a_2}{a_1}\right)^3
\left(\frac{425-121\epsilon_{\rm min}}{\sin^2i_{\rm crit}\epsilon_{\rm min}^3}\right)
(1-e_2^2)^{3/2}.
\label{sing}
\eeq
Using the fact that (Wen 2003)
\beq
\cos i_{\rm crit}=\cos i_0\left(\frac{\epsilon_0}{\epsilon_{\rm min}}\right)^{1/2}
+\frac{a_1}{2\beta_0}\frac{1}{\sqrt{a_1\epsilon_{\rm min}}}\left(\epsilon_0-\epsilon_{\rm min}\right),
\eeq
where $\beta_0=(m_2/m_0m_1)[(m_0+m_1)^3a_2/((m_0+m_1+m_2)(1-e_2^2))]^{1/2}$,
allows one to write $\sin(2g_{1,\,\rm crit})$ (equation \ref{sing}) 
in terms of  just $\epsilon_{\rm min}$ and the initial parameters of the 
system.  In calculating $g_{1,\,\rm crit}$ and $\cos i_{\rm crit}$, I assume
that $a_1$, $a_2$, and $e_2$ are unchanged from their initial values.
Solving the implicit equation $\Delta W= W(\epsilon_0,g_{\rm 1,\,0})-W(\epsilon_{\rm min},
g_{1\,\rm crit})=0$ for $\epsilon_{\rm min}$, then allows for an accurate
estimate of the merger time (MH02):
\beq
t_{\rm merge}=t_{\rm GW}(a_{1,0},\epsilon_{\rm min})\epsilon_{\rm min}^{-1/2},
\label{est}
\eeq
where here $t_{\rm GW}$ is computed from the formalism of Peters (1964).
In practice, I solve the equation $\Delta W=0$ using Newton-Raphson 
iteration with an initial guess for $\epsilon_{\rm min}$ using the 
approximate expressions from Wen (2003) and MH02.

\begin{figure*}
\centerline{\includegraphics[width=8cm]{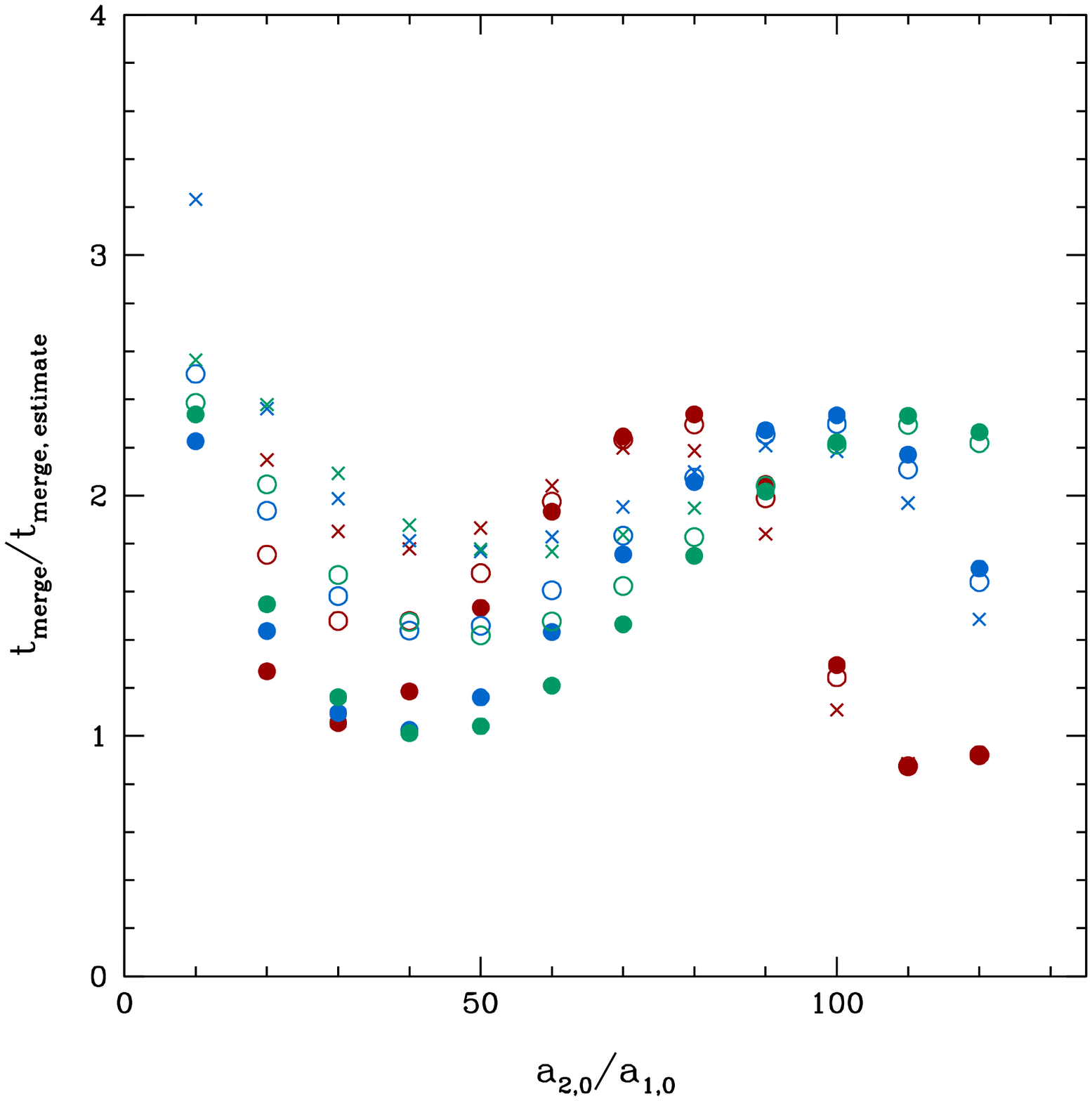}\includegraphics[width=8cm]{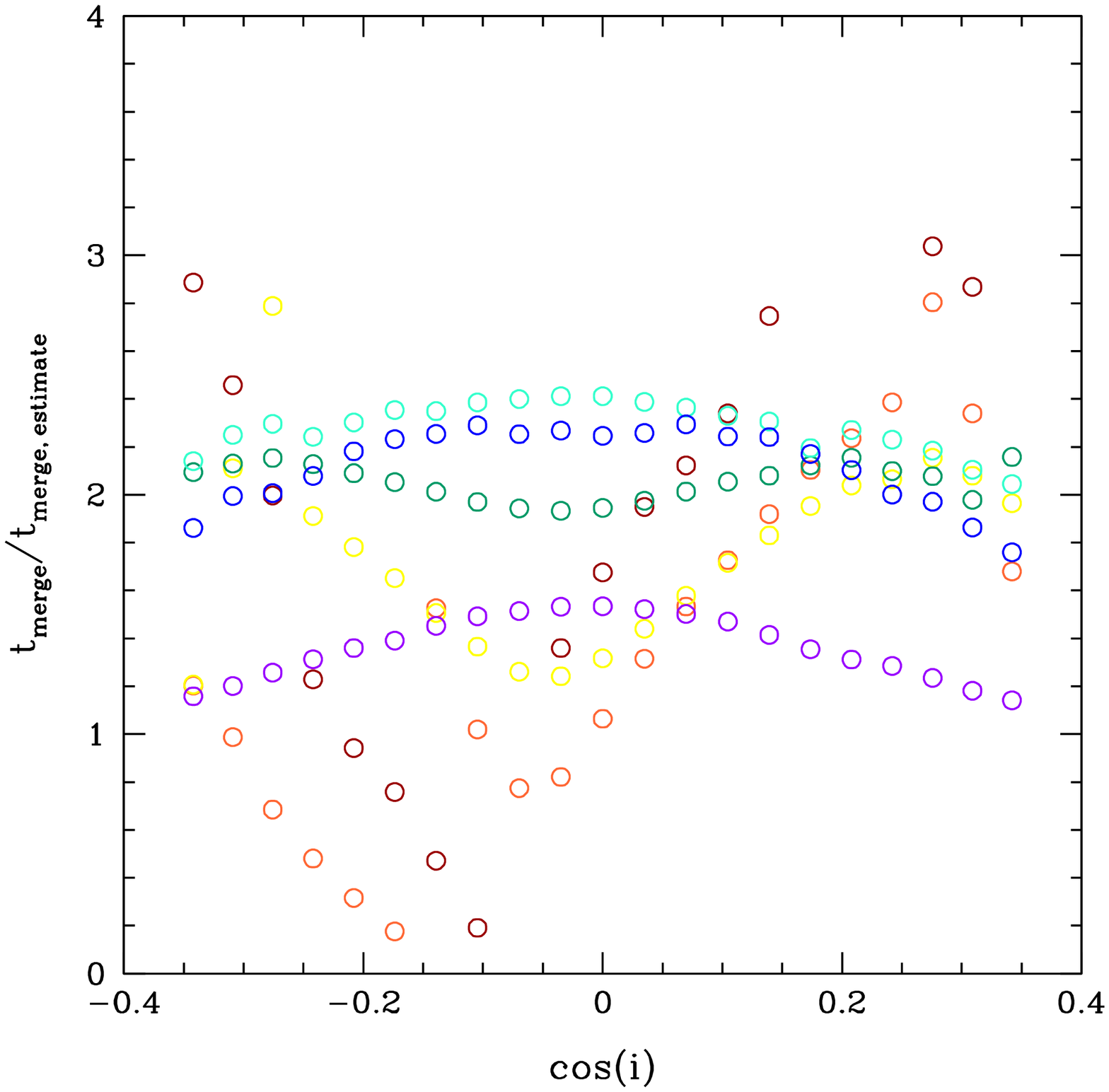}}
\caption{{\it Left panel:} Comparison between $t_{\rm merge}$ calculated
by solving the full system of equations and the estimate
of $t_{\rm merge}$ given in equation (\ref{est})
as a function of $a_{2,0}/a_{1,0}$, for $a_{1,0}=0.03$ (red), 0.06 (blue), and 0.09 (green),  
for $i=89^\circ$ (filled circles),  $85^\circ$ (open circles), and $80^\circ$ (crosses).  
The estimated merger timescale typically underpredicts $t_{\rm merge}$ by a factor 
of $\sim1.5-2.5$, with systematic offset that depends on $i$
and $a_2/a_1$. {\it Right panel:} Same comparison, but as a function of $\cos(i)$,
for $a_1=0.01$\,AU, and $a_2/a_1=10$ (red), 20 (orange), 30 (yellow), 
40 (green), 50 (turquoise), 60 (blue),  and 70 (violet).  
Compare with right panel of Fig.~\ref{fig:wd}.
}
\label{fig:comp}
\end{figure*}

A comparison between the numerically calculated value of $t_{\rm merge}$
and this approximate scheme is shown in Figure \ref{fig:comp}.
Typically, $t_{\rm merge}$ as calculated from 
equation (\ref{est}) is underestimated by a factor of
$\sim1.5-2$.  There are systematic deviations
as a function of both $i$ and $a_2/a_1$.  These deviations
are not due to the fact that the numerical solution is 
to the octopole-level equations, whereas equation (\ref{est})
is only at quadrupole order, since for $m_0=m_1$ the 
equations reduce from octopole to quadrupole and 
this hypothesis can be explicitly tested.  
The right panel shows that for retrograde
orbits and small $a_2/a_1 \leq20$ the estimate of equation (\ref{est})
can overpredict $t_{\rm merge}$ by a factor of $\sim10$.  Note, 
however, that it is precisely in these cases where tidal effects
(neglected here; \S\ref{section:method}) 
are expected to be most important to the evolution.
 Future work should address these systematic differences 
between the $t_{\rm merge}$ and the estimate of 
equation (\ref{est}).  
For the purposes of a quick and
fairly accurate estimate of $t_{\rm merge}$ (as in  Figs.~\ref{fig:ain} and \ref{fig:data}), Figure
\ref{fig:comp} shows that  equation (\ref{est}) is sufficient.

%%%%%%%%%%%%%%%%%%%%%%%%%
%%%%%%%%%%%%%%%%%%%%%%%%% 

\end{appendix}


\begin{thebibliography}{}
\bibitem[Aubourg et al.(2008)]{2008A&A...492..631A} 
Aubourg, {\'E}., Tojeiro, R., Jimenez, R., Heavens, A., 
Strauss, M.~A., \& Spergel, D.~N.\ 2008, \aap, 492, 631 

\bibitem[Badenes et al.(2009)]{2009ApJ...707..971B} Badenes, C., Mullally, 
F., Thompson, S.~E., \& Lupton, R.~H.\ 2009, \apj, 707, 971 

\bibitem[Bhattacharya \& van den Heuvel(1991)]{1991PhR...203....1B} 
Bhattacharya, D., \& van den Heuvel, E.~P.~J.\ 1991, \physrep, 203, 1 


\bibitem[Belczy{\'n}ski \& Kalogera(2001)]{2001ApJ...550L.183B} 
Belczy{\'n}ski, K., \& Kalogera, V.\ 2001, \apjl, 550, L183 

\bibitem[Benacquista(2001)]{2001AIPC..586..793B} Benacquista, M.\ 2001, 
20th Texas Symposium on relativistic astrophysics, 586, 793 

\bibitem[Berger(2009)]{2009ApJ...690..231B} Berger, E.\ 2009, \apj, 690, 
231 

\bibitem[Bildsten et al.(2007)]{2007ApJ...662L..95B} Bildsten, L., Shen, 
K.~J., Weinberg, N.~N., \& Nelemans, G.\ 2007, \apjl, 662, L95 


\bibitem[Blaes et al.(2002)]{2002ApJ...578..775B} 
Blaes, O., Lee, M.~H., \& Socrates, A.\ 2002, \apj, 578, 775 (BLS02)

\bibitem[Brandt et al.(2010)]{2010AJ....140..804B} Brandt, T.~D., Tojeiro, 
R., Aubourg, {\'E}., Heavens, A., Jimenez, R., 
\& Strauss, M.~A.\ 2010, \aj, 140, 804 

\bibitem[Brown et al.(2010)]{2010ApJ...723.1072B} Brown, W.~R., Kilic, M., 
Allende Prieto, C., \& Kenyon, S.~J.\ 2010, \apj, 723, 1072 

\bibitem[Champion et al.(2008)]{2008Sci...320.1309C} Champion, D.~J., et 
al.\ 2008, Science, 320, 1309 

\bibitem[Eggleton \& Kiseleva-Eggleton(2001)]{2001ApJ...562.1012E} 
Eggleton, P.~P., \& Kiseleva-Eggleton, L.\ 2001, \apj, 562, 1012 

\bibitem[Eggleton \& Kiseleva(1995)]{1995ApJ...455..640E} 
Eggleton, P., \& Kiseleva, L.\ 1995, \apj, 455, 640 

\bibitem[Evans et al.(2005)]{2005AJ....130..789E} Evans, N.~R., Carpenter, 
K.~G., Robinson, R., Kienzle, F., \& Dekas, A.~E.\ 2005, \aj, 130, 789 

\bibitem[Exter et al.(2010)]{2010AJ....140.1414E} Exter, K., Bond, H.~E., 
Stassun, K.~G., Smalley, B., Maxted, P.~F.~L., 
\& Pollacco, D.~L.\ 2010, \aj, 140, 1414 

\bibitem[Fabrycky \& Tremaine(2007)]{2007ApJ...669.1298F} 
Fabrycky, D., \& Tremaine, S.\ 2007, \apj, 669, 1298 (FT07)

\bibitem[Farmer \& Phinney(2003)]{2003MNRAS.346.1197F} 
Farmer, A.~J., \& Phinney, E.~S.\ 2003, \mnras, 346, 1197 

\bibitem[Ford et al.(2000)]{2000ApJ...535..385F} 
Ford, E.~B., Kozinsky, B., \& Rasio, F.~A.\ 2000, \apj, 535, 385

\bibitem[Ford et al.(2004)]{2004ApJ...605..966F} 
Ford, E.~B., Kozinsky, B., \& Rasio, F.~A.\ 2004, \apj, 605, 966 

\bibitem[Fregeau et al.(2004)]{2004MNRAS.352....1F} Fregeau, J.~M., Cheung, 
P., Portegies Zwart, S.~F., \& Rasio, F.~A.\ 2004, \mnras, 352, 1 

\bibitem[Fregeau et al.(2009)]{2009ApJ...707.1533F} Fregeau, J.~M., 
Ivanova, N., \& Rasio, F.~A.\ 2009, \apj, 707, 1533 

\bibitem[Freire et al.(2011)]{2011MNRAS.412.2763F} Freire, P.~C.~C., et 
al.\ 2011, \mnras, 412, 2763 

\bibitem[Fryer et al.(1999)]{1999ApJ...526..152F} 
Fryer, C.~L., Woosley, S.~E., \& Hartmann, D.~H.\ 1999b, \apj, 526, 152 

\bibitem[Fryer \& Woosley(1998)]{1998ApJ...502L...9F} 
Fryer, C.~L., \& Woosley, S.~E.\ 1998, \apjl, 502, L9 

\bibitem[Fryer et al.(1999)]{1999ApJ...520..650F} Fryer, C.~L., Woosley, 
S.~E., Herant, M., \& Davies, M.~B.\ 1999a, \apj, 520, 650 

\bibitem[Gould(2011)]{2011ApJ...729L..23G} Gould, A.\ 2011, \apjl, 729, L23 

\bibitem[Hills(1983)]{1983ApJ...267..322H} Hills, J.~G.\ 1983, \apj, 267, 
322 

\bibitem[Holman et al.(1997)]{1997Natur.386..254H} 
Holman, M., Touma, J., \& Tremaine, S.\ 1997, \nat, 386, 254 

\bibitem[Horiuchi et al.(2009)]{2009PhRvD..79h3013H} 
Horiuchi, S., Beacom, J.~F., \& Dwek, E.\ 2009, \prd, 79, 083013 

\bibitem[Horiuchi \& Beacom(2010)]{2010ApJ...723..329H} 
Horiuchi, S., \& Beacom, J.~F.\ 2010, \apj, 723, 329 

\bibitem[Iben 
\& Tutukov(1984)]{1984ApJS...54..335I} 
Iben, I., Jr., \& Tutukov, A.~V.\ 1984, \apjs, 54, 335 

\bibitem[Iben 
\& Tutukov(1985)]{1985ApJS...58..661I} 
Iben, I., Jr., \& Tutukov, A.~V.\ 1985, \apjs, 58, 661 

\bibitem[Iben 
\& Tutukov(1987)]{1987ApJ...313..727I} 
Iben, I., Jr., \& Tutukov, A.~V.\ 1987, \apj, 313, 727 

\bibitem[Iben \& Livio(1993)]{1993PASP..105.1373I} 
Iben, I., Jr., \& Livio, M.\ 1993, \pasp, 105, 1373 

\bibitem[Iben \& Tutukov(1999)]{1999ApJ...511..324I} 
Iben, I., Jr., \& Tutukov, A.~V.\ 1999, \apj, 511, 324 

\bibitem[Innanen et al.(1997)]{1997AJ....113.1915I} Innanen, K.~A., Zheng, 
J.~Q., Mikkola, S., \& Valtonen, M.~J.\ 1997, \aj, 113, 1915 

\bibitem[Ivanova(2008)]{2008msah.conf..101I} Ivanova, N.\ 2008, Multiple 
Stars Across the H-R Diagram, 101 

\bibitem[Ivanova et al.(2006)]{2006MNRAS.372.1043I} Ivanova, N., Heinke, 
C.~O., Rasio, F.~A., Taam, R.~E., Belczynski, K., 
\& Fregeau, J.\ 2006, \mnras, 372, 1043 

\bibitem[Ivanova et al.(2008)]{2008MNRAS.386..553I} Ivanova, N., Heinke, 
C.~O., Rasio, F.~A., Belczynski, K., 
\& Fregeau, J.~M.\ 2008, \mnras, 386, 553 

\bibitem[Ivanova et al.(2010)]{2010ApJ...717..948I} Ivanova, N., 
Chaichenets, S., Fregeau, J., Heinke, C.~O., Lombardi, J.~C., 
\& Woods, T.~E.\ 2010, \apj, 717, 948 



\bibitem[Janka et al.(1999)]{1999ApJ...527L..39J} Janka, H.-T., Eberl, T., 
Ruffert, M., \& Fryer, C.~L.\ 1999, \apjl, 527, L39 

\bibitem[Kalirai et al.(2008)]{2008ApJ...676..594K} Kalirai, J.~S., Hansen, 
B.~M.~S., Kelson, D.~D., Reitzel, D.~B., Rich, R.~M., 
\& Richer, H.~B.\ 2008, \apj, 676, 594 



\bibitem[Kalogera(1996)]{1996ApJ...471..352K} Kalogera, V.\ 1996, \apj, 
471, 352 

\bibitem[Kalogera et al.(2001)]{2001ApJ...556..340K} Kalogera, V., Narayan, 
R., Spergel, D.~N., \& Taylor, J.~H.\ 2001, \apj, 556, 340 

\bibitem[Kalogera et al.(2004a)]{2004ApJ...601L.179K} Kalogera, V., et al.\ 
2004a, \apjl, 601, L179 

\bibitem[Kalogera et al.(2004b)]{2004ApJ...614L.137K} Kalogera, V., et al.\ 
2004b, \apjl, 614, L137 


\bibitem[Kasen(2010)]{2010ApJ...708.1025K} Kasen, D.\ 2010, \apj, 708, 1025 

\bibitem[Kerzendorf et al.(2009)]{2009ApJ...701.1665K} Kerzendorf, W.~E., 
Schmidt, B.~P., Asplund, M., Nomoto, K., Podsiadlowski, P., Frebel, A., 
Fesen, R.~A., \& Yong, D.\ 2009, \apj, 701, 1665 

\bibitem[Kilic et al.(2010a)]{2010ApJ...716..122K} Kilic, M., Brown, W.~R., 
Allende Prieto, C., Kenyon, S.~J., \& Panei, J.~A.\ 2010a, \apj, 716, 122 

\bibitem[Kilic et al.(2010b)]{2010ApJ...721L.158K} Kilic, M., Allende 
Prieto, C., Brown, W.~R., Ag{\"u}eros, M.~A., Kenyon, S.~J., 
\& Camilo, F.\ 2010b, \apjl, 721, L158 

\bibitem[Kiseleva et al.(1998)]{1998MNRAS.300..292K} Kiseleva, L.~G., 
Eggleton, P.~P., \& Mikkola, S.\ 1998, \mnras, 300, 292 

\bibitem[Kochanek(2009)]{2009ApJ...707.1578K} 
Kochanek, C.~S.\ 2009, \apj, 707, 1578 

\bibitem[Kozai(1962)]{1962AJ.....67..591K} Kozai, Y.\ 1962, \aj, 67, 591 

\bibitem[Krymolowski 
\& Mazeh(1999)]{1999MNRAS.304..720K} Krymolowski, Y., \& Mazeh, T.\ 1999, \mnras, 304, 720 

\bibitem[Kulkarni 
\& van Kerkwijk(2010)]{2010ApJ...719.1123K} Kulkarni, S.~R., \& van Kerkwijk, M.~H.\ 2010, \apj, 719, 1123 


\bibitem[Lada(2006)]{2006ApJ...640L..63L} Lada, C.~J.\ 2006, \apjl, 640, L63 

\bibitem[MacFadyen et al.(2005)]{2005astro.ph.10192M} MacFadyen, A.~I., 
Ramirez-Ruiz, E., \& Zhang, W.\ 2005, arXiv:astro-ph/0510192 

\bibitem[Mannucci et al.(2006)]{2006MNRAS.370..773M} Mannucci, F., Della 
Valle, M., \& Panagia, N.\ 2006, \mnras, 370, 773 

\bibitem[Maoz(2010)]{2010arXiv1011.1014M} Maoz, D.\ 2010, arXiv:1011.1014 

\bibitem[Maoz \& Badenes(2010)]{2010MNRAS.tmp..968M} 
Maoz, D., \& Badenes, C.\ 2010, \mnras, 968 

\bibitem[Maoz et al.(2010)]{2010ApJ...722.1879M} Maoz, D., Sharon, K., 
\& Gal-Yam, A.\ 2010, \apj, 722, 1879 

\bibitem[Maoz et al.(2011)]{2011MNRAS.412.1508M} Maoz, D., Mannucci, F., 
Li, W., Filippenko, A.~V., Valle, M.~D., 
\& Panagia, N.\ 2011, \mnras, 412, 1508 

\bibitem[Marchal(1990)]{1990tbp..book.....M} Marchal, C.\ 1990, Studies in 
Astronautics, Studies in Aeronautics, 4.~Amsterdam: Elsevier, 1990

\bibitem[Mardling \& Aarseth(2001)]{2001MNRAS.321..398M} 
Mardling, R.~A., \& Aarseth, S.~J.\ 2001, \mnras, 321, 398 

\bibitem[Marietta et al.(2000)]{2000ApJS..128..615M} Marietta, E., Burrows, 
A., \& Fryxell, B.\ 2000, \apjs, 128, 615 

\bibitem[Mazeh \& Shaham(1979)]{1979A&A....77..145M} 
Mazeh, T., \& Shaham, J.\ 1979, \aap, 77, 145 

\bibitem[Metzger(2011)]{2011arXiv1105.6096M} 
Metzger, B.~D.\ 2011, arXiv:1105.6096 

\bibitem[Miller \& Hamilton(2002)]{2002ApJ...576..894M} 
Miller, M.~C., \& Hamilton, D.~P.\ 2002, \apj, 576, 894 (MH02)

\bibitem[Mullally et al.(2009)]{2009ApJ...707L..51M} Mullally, F., Badenes, 
C., Thompson, S.~E., \& Lupton, R.\ 2009, \apjl, 707, L51 

\bibitem[Nakar(2007)]{2007PhR...442..166N} Nakar, E.\ 2007, \physrep, 442, 166 

\bibitem[Napiwotzki et al.(2001)]{2001AN....322..411N} Napiwotzki, R., et 
al.\ 2001, Astronomische Nachrichten, 322, 411 

\bibitem[Nelemans et al.(2001)]{2001A&A...368..939N} 
Nelemans, G., Portegies Zwart, S.~F., Verbunt, F., \& Yungelson, L.~R.\ 2001, \aap, 368, 939 

\bibitem[Nelemans et 
al.(2005)]{2005A&A...440.1087N} Nelemans, G., et al.\ 2005, \aap, 440, 1087 



\bibitem[O'Shaughnessy et al.(2008)]{2008ApJ...675..566O} O'Shaughnessy, 
R., Belczynski, K., \& Kalogera, V.\ 2008, \apj, 675, 566 

\bibitem[Pakmor et al.(2010)]{2010Natur.463...61P} 
Pakmor, R., Kromer, M., 
R{\"o}pke, F.~K., Sim, S.~A., Ruiter, A.~J., 
\& Hillebrandt, W.\ 2010, \nat, 463, 61 

\bibitem[Perets \& Fabrycky(2009)]{2009ApJ...697.1048P} 
Perets, H.~B., \& Fabrycky, D.~C.\ 2009, \apj, 697, 1048 

\bibitem[Peters(1964)]{peters}Peters, P.\ C.\ 1964, Phys.\ Rev.\ B, 136, 1224

\bibitem[Pinsonneault \& Stanek(2006)]{2006ApJ...639L..67P} 
Pinsonneault, M.~H., \& Stanek, K.~Z.\ 2006, \apjl, 639, L67 

\bibitem[Press et al.(1992)]{1992nrfa.book.....P} 
Press, W.~H., Teukolsky, S.~A., Vetterling, W.~T., 
\& Flannery, B.~P.\ 1992, Cambridge: University Press, c1992, 2nd ed.,  

\bibitem[Pribulla \& Rucinski(2006)]{2006AJ....131.2986P} 
Pribulla, T., \& Rucinski, S.~M.\ 2006, \aj, 131, 2986 

\bibitem[Pritchet et al.(2008)]{2008ApJ...683L..25P} Pritchet, C.~J., 
Howell, D.~A., \& Sullivan, M.\ 2008, \apjl, 683, L25 

\bibitem[Raghavan et al.(2010)]{2010ApJS..190....1R} Raghavan, D., et al.\ 2010, \apjs, 190, 1 

\bibitem[Raskin et al.(2010)]{2010ApJ...724..111R} Raskin, C., Scannapieco, 
E., Rockefeller, G., Fryer, C., Diehl, S., 
\& Timmes, F.~X.\ 2010, \apj, 724, 111 

\bibitem[Rosswog et al.(2009)]{2009ApJ...705L.128R} Rosswog, S., Kasen, D., 
Guillochon, J., \& Ramirez-Ruiz, E.\ 2009, \apjl, 705, L128 

\bibitem[Ruiter et al.(2009)]{2009ApJ...699.2026R} Ruiter, A.~J., 
Belczynski, K., \& Fryer, C.\ 2009, \apj, 699, 2026 


\bibitem[Ruffert \& Janka(1999)]{1999A&A...344..573R} 
Ruffert, M., \& Janka, H.-T.\ 1999, \aap, 344, 573 

\bibitem[Saio \& Nomoto(1985)]{1985A&A...150L..21S} Saio, H., \& Nomoto, K.\ 1985, \aap, 150, L21 

\bibitem[Scannapieco \& Bildsten(2005)]{2005ApJ...629L..85S} 
Scannapieco, E., \& Bildsten, L.\ 2005, \apjl, 629, L85 

\bibitem[Stairs(2004)]{2004Sci...304..547S} Stairs, I.~H.\ 2004, Science, 
304, 547 

\bibitem[Tokovinin et 
al.(2006)]{2006A&A...450..681T} Tokovinin, A., Thomas, S., Sterzik, M., \& Udry, S.\ 2006, \aap, 450, 681 

\bibitem[Totani et al.(2008)]{2008PASJ...60.1327T} Totani, T., Morokuma, 
T., Oda, T., Doi, M., \& Yasuda, N.\ 2008, \pasj, 60, 1327 


\bibitem[van Kerkwijk et al.(2010)]{2010ApJ...722L.157V} van Kerkwijk, 
M.~H., Chang, P., \& Justham, S.\ 2010, \apjl, 722, L157 

\bibitem[Warner(1995)]{1995Ap&SS.225..249W} Warner, B.\ 1995, \apss, 225, 249 

\bibitem[Webbink(1984)]{1984ApJ...277..355W} 
Webbink, R.~F.\ 1984, \apj, 277, 355 

\bibitem[Wen(2003)]{2003ApJ...598..419W} 
Wen, L.\ 2003, \apj, 598, 419 

\bibitem[Wheeler et al.(1975)]{1975ApJ...200..145W} Wheeler, J.~C., Lecar, 
M., \& McKee, C.~F.\ 1975, \apj, 200, 145 

\bibitem[Whelan 
\& Iben(1973)]{1973ApJ...186.1007W} Whelan, J., \& Iben, I., Jr.\ 1973, \apj, 186, 1007 

\bibitem[Willems et al.(2007)]{2007ApJ...665L..59W} Willems, B., Kalogera, 
V., Vecchio, A., Ivanova, N., Rasio, F.~A., Fregeau, J.~M., 
\& Belczynski, K.\ 2007, \apjl, 665, L59 

\bibitem[Wu \& Murray(2003)]{2003ApJ...589..605W} 
Wu, Y., \& Murray, N.\ 2003, \apj, 589, 605 

\bibitem[Wu et al.(2007)]{2007ApJ...670..820W} Wu, Y., Murray, N.~W., 
\& Ramsahai, J.~M.\ 2007, \apj, 670, 820 



\end{thebibliography}
\end{document}